\documentclass[twocolumn]{aastex631}
\usepackage{amssymb}
\usepackage{amsmath}
\usepackage{mathtools}
\usepackage{xcolor}
\usepackage{array}
\usepackage{tabularx}
\usepackage{multirow}
\usepackage{graphicx}
\shorttitle{Extragalactic AME detection in NGC 2903 using KVN}
\shortauthors{POOJON et al.}
\graphicspath{{./}{figures/}}
\begin{document}

\title{Detection of extragalactic anomalous microwave emission in NGC 2903 \\ using KVN single-dish observations}

\author{Panomporn Poojon}
\affiliation{Department of Astronomy, Yonsei University, 50 Yonsei-ro, Seodaemun-gu, Seoul 03722, Republic of Korea}
\email{poojon.p@gmail.com, achung@yonsei.ac.kr,\\ thiemhoang@kasi.re.kr}


\author[0000-0003-1440-8552]{Aeree Chung}
\affiliation{Department of Astronomy, Yonsei University, 50 Yonsei-ro, Seodaemun-gu, Seoul 03722, Republic of Korea}

\author[0000-0003-2017-0982]{Thiem Hoang}
\affiliation{Korea Astronomy and Space Science Institute, 776 Daedeokdae-ro, Yuseong-gu, Daejeon 34055, Republic of Korea}
\affiliation{Department of Astronomy and Space Science, University of Science and Technology, 217 Gajeong-ro, Yuseong-gu, Daejeon, 34113, Republic of Korea}

\author[0000-0002-3744-6714]{Junhyun Baek}
\affiliation{Department of Astronomy, Yonsei University, 50 Yonsei-ro, Seodaemun-gu, Seoul 03722, Republic of Korea}
\affiliation{Korea Astronomy and Space Science Institute, 776 Daedeokdae-ro, Yuseong-gu, Daejeon 34055, Republic of Korea}

\author{Hiroyuki Nakanishi}
\affiliation{Graduate School of Science and Engineering, Kagoshima University, 1-21-35 Korimoto, Kagoshima, Kagoshima 890-0065, Japan}

\author[0000-0003-1659-095X]{Tomoya Hirota}
\affiliation{Mizusawa VLBI Observatory, National Astronomical Observatory of Japan,
2-12 Hoshiga-oka, Mizusawa, Oshu-shi, Iwate 023-0861, Japan}
\affiliation{The Graduate University for Advanced Studies, SOKENDAI,
2-21-1 Osawa, Mitaka, Tokyo 181-8588, Japan}

\author[0000-0002-9390-9672]{Chao-Wei Tsai}
\affiliation{National Astronomical Observatories, Chinese Academy of Sciences, 20A Datun Road, Beijing 100101, China}
\affiliation{Institute for Frontiers in Astronomy and Astrophysics, Beijing Normal University, Beijing 102206, China}
\affiliation{School of Astronomy and Space Science, University of Chinese Academy of Sciences, Beijing 100049, China}



\begin{abstract}
We present the results of the single-dish observations using the Korean VLBI Network to search for anomalous microwave emission (AME) in nearby galaxies. The targets were selected from MApping the dense molecular gAs in the strongest star-formiNg Galaxies (MALATANG), a legacy survey project of the James Clerk Maxwell Telescope. The MALATANG galaxies are good representatives of local galaxies with enhanced nuclear activity associated with star formation and/or AGN, providing IR-bright galaxy samples; thus, they are good candidates for AME hosts. Combining with the ancillary data, we investigated the radio-IR spectral energy distribution (SED), while searching for the AME signals in five galaxies. The AME in NGC 2903 was well detected at a significant confidence level, whereas that in NGC 2146 and M82 was marginal. NGC 1068 and Arp 299 indicated no significant hints, and we provided the upper limits for the AME. The best-fit SED exhibited local peaks of the AME components at higher frequencies and with stronger peak fluxes than those in the previous studies. This suggested the origin of AME being denser environments such as molecular clouds or photodissociation regions rather than warm neutral/ionized medium as commonly suggested by previous studies. Further, our AME-detected targets were observed to exhibit higher specific star-formation rates than the other extragalactic AME hosts. Furthermore, AME favored starburst galaxies among our sample rather than AGN hosts. Consequently, this might imply that AGNs are excessively harsh environments for tiny dust to survive.
\end{abstract}



\section{INTRODUCTION} \label{sec:intro}
Anomalous microwave emission (AME) is a diffuse galactic emission component in the frequency range of $10-100$ GHz. It was first discovered as a bi-product from the observations of the Cosmic Background Explorer (COBE) \citep{1996ApJ...464L...5K,1997ApJ...486L..23L}. The original explanation for AME is that they are emissions from rapidly spinning polycyclic aromatic hydrocarbons (PAHs) with permanent electric dipole moments, which is often referred to as a spinning dust mechanism \citep{1998ApJ...494L..19D, 1998ApJ...508..157D} (hereafter DL98). However, there are studies refuting this hypothesis by reporting a fainter correlation between AME and PAH emission \citep{Hensley_2016, Hensley_2017}. \citet{1999ApJ...512..740D} has also suggested microwave emission arising from thermal fluctuations of magnetic dipole moments within ferro/ferri-magnetic grains as a potential mechanism of AME. Nevertheless, subsequent observations have supported the spinning dust mechanism as the leading mechanism of AME \citep{Oliveira_2002, 2004ApJ...617..350F}.

\citet{AliHaimoud_2009} improved the original spinning dust model of DL98. They computed the angular velocity distribution of grains by applying the Fokker-Planck equation instead of the Maxwellian thermal distribution used in DL98. Consequently, more comprehensive models have been introduced by implementing important physical effects such as grain rotation around its non-principal axis \citep{Hoang_2011, Silsbee_2011}, transient spin-up of grains \citep{Hoang_2011}, and irregular grain shape \citep{Hoang_2011}. In addition to spinning PAHs, recent studies have suggested that spinning very small grains (VSGs) of radius $\lesssim 10$ nm, including nanosilicates \citep{Hoang_2016_2} and iron nanoparticles \citep{Hoang_2016}, with permanent electric/magnetic dipole moments are potential sources of spinning dust \citep{Hoang_2016, Hensley_2017}.\footnote{Throughout this paper, VSGs, and nanoparticles are used interchangeably.} However, the exact mechanism of AME and its carrier (PAHs, nanosilicate, and nanoiron) remains debatable. Therefore, AME observations are crucial to understand the nature and the physics of spinning dust.

To date, AME has been observed in numerous regions inside the Milky Way, such as the Perseus molecular cloud \citep{2005ApJ...624L..89W, 2011A&A...536A..20P}, $\rho$ Ophiuchi molecular clouds associated with the photodissociation region \citep{Casassus_2008, 2011A&A...536A..20P}, and the HII region - RCW175 \citep{Tibbs_2012, Dickinson_2009,Battistelli_2015}. The properties of Galactic AME from different regions do not vary that much with a common peak frequency of approximately 30 GHz.

AME observations in other galaxies of various properties are also important for a comprehensive understanding of the nature of AME and its implication for probing the properties of nanodust. However, till date, AME has been reported only in a few extragalactic systems. \citet{Murphy_2010} first detected AME associated with a star-forming region in a nearby galaxy, NGC 6946, which has been confirmed by \citet{Hensley_2015}. AME was also found in the Large Magellanic Cloud (LMC) and the Small Magellanic Cloud \citep{2011A&A...536A..17P}. \citet{2015A&A...582A..28P} reported a marginal detection of AME in the Andromeda galaxy (M31). However, more significant evidence supporting the presence of AME in M31 was subsequently confirmed by \citet{Battistelli_2019}. A recent study by \citet{2023MNRAS.523.3471H} reported the marginal detection of AME in M31, suggesting that it may not be uniformly distributed throughout the entire galaxy. Conversely, the study by \citet{2023arXiv230508547F} presented a contrasting view, strongly advocating for the presence of AME in M31. In addition, \citet{Tibbs_2018} reported an AME-like source in a small spiral galaxy in the Local Group, M33, in its global spectral energy distribution (SED). Although AME has been probed in several other galaxies, including NGC 253, NGC 4945, and M82 \citep{Peel_2011} and NGC 3627, NGC 4254, NGC 4736, and NGC 5055 \citep{Bianchi_2022}, only the upper limits have been obtained in those nearby galaxies.

This study reported our search for evidence of AME among a sample of IR-bright nearby galaxies with enhanced nuclear activity associated with star formation and/or active galactic nuclei (AGN). We verified the presence of AME by performing SED fitting analysis on the radio-IR data and investigated the correlation of AME properties with galactic properties.

The remainder of this paper is organized as follows. In Section \ref{sec:sam_kvn}, we describe the sample selection and the observations related to AME in extragalactic systems using the Korean VLBI Network (KVN). The details of ancillary data and the photometry are presented in Section \ref{sec:anc_data}. We describe the SED fitting procedure and present the best-fitted parameters for observed SEDs in section \ref{sec:sedfit}. Section \ref{sec:result} presents the AME properties, and in Section \ref{sec:dis}, the properties of AME and galaxy are discussed. Finally, we summarize and conclude the results in Section \ref{sec:conc}.

\section{SAMPLE AND OBSERVATION}  \label{sec:sam_kvn}
\subsection{Sample selection and KVN observation}
The sample used in this study was selected from the Mapping the dense molecular gas in the strongest star-forming galaxies (MALATANG) study. MALATANG is a James Clerk Maxwell Telescope (JCMT) mapping legacy survey of the nearest 23 IR-brightest galaxies in dense molecular lines such as HCN and HCO+ (4-3) \citep{Tan_2018, Jiang_2020}. As good representatives of local galaxies with enhanced nuclear activity associated with intensive star formation and/or AGN, the MALATANG sample provides ideal targets for the AME search. AME has been reported in one of the MALATANG galaxies, that is, the star-forming region of NGC 6946 by \citet{Murphy_2010}. In addition, comprehensive ancillary data are already available for several MALATANG galaxies, rendering it easier to probe the correlation.

The KVN is a very long baseline interferometry (VLBI) network comprising three 21 m radio antennas located at the Yonsei, Ulsan, and Tamna sites that provide the longest baseline of $\sim$ 500 km when combined. While its VLBI resolution is excessively high to detect the expected flux densities of most MALATANG galaxies within those frequencies, even in the presence of an AME bump, it can be operated individually or as single-dishes. The KVN is a good facility for this study as it is equipped with bands at four frequencies of 22, 43, 86, and 129 GHz. Thus, the four bands at frequencies of 22, 43, 86, and 129 GHz available on the Korean VLBI Network dishes render this facility as ideal for investigating the AME bump.

We conducted KVN single-dish, multi-band observations during 2017-2020. The beam sizes are $\approx$ 130”, 65”, 32”, and 23” at 22, 43, 86, and 129 GHz, respectively. We used the cross-scan mode along the RA-Dec direction to safely escape the target and secure the reference sky. 

\begin{figure*}[]
  \centering
  \begin{tabular}{ccccc}
    \includegraphics[width=.2\linewidth]{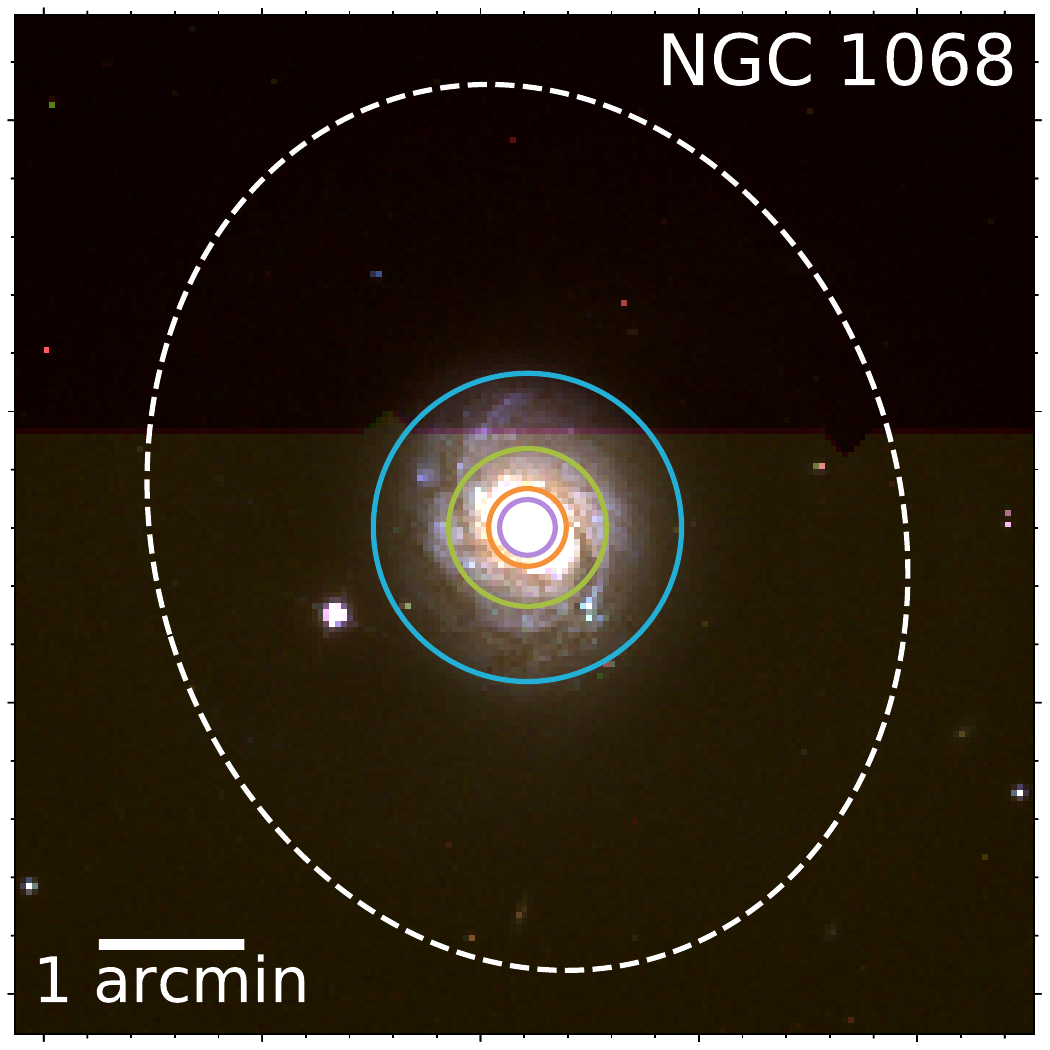} & \hspace{-0.5cm}
	\includegraphics[width=.2\linewidth]{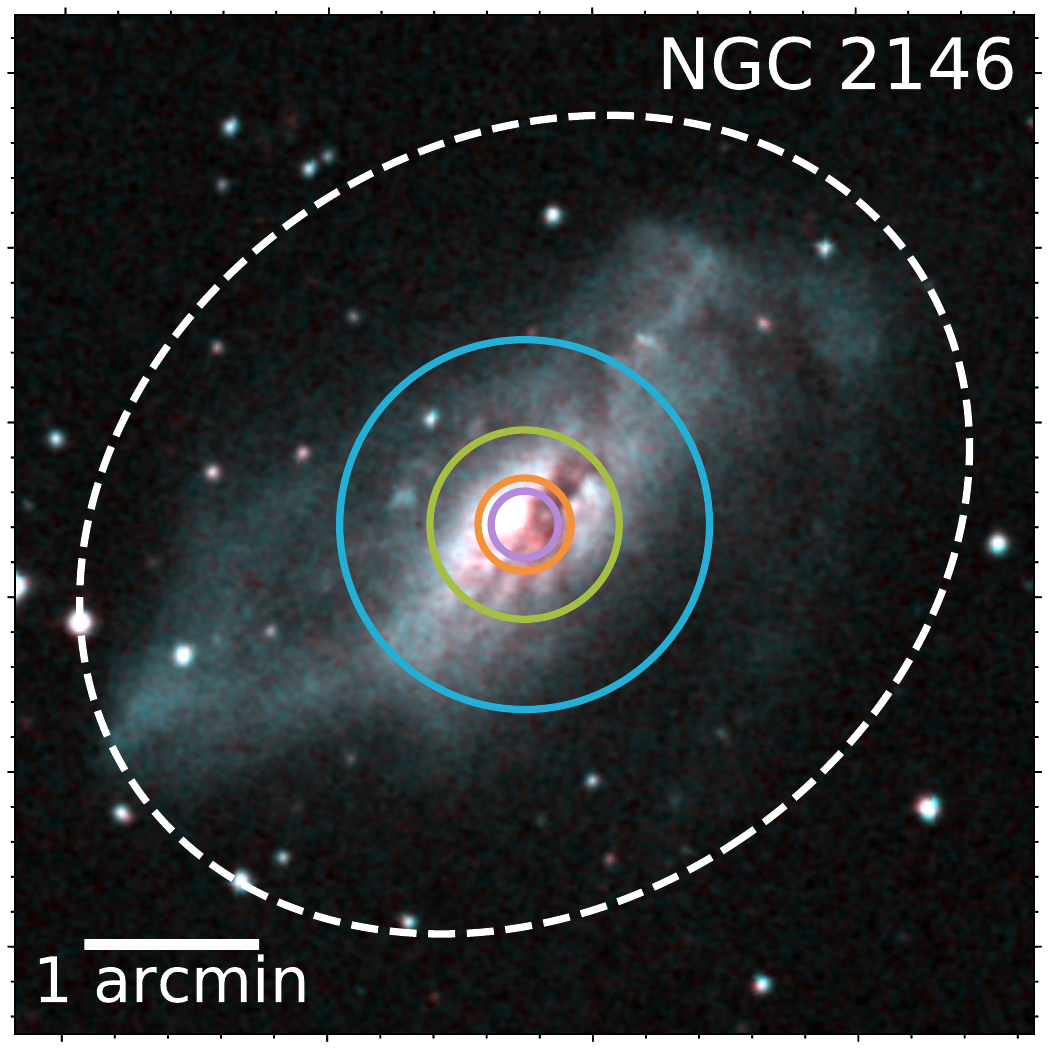} & \hspace{-0.5cm}
    \includegraphics[width=.2\linewidth]{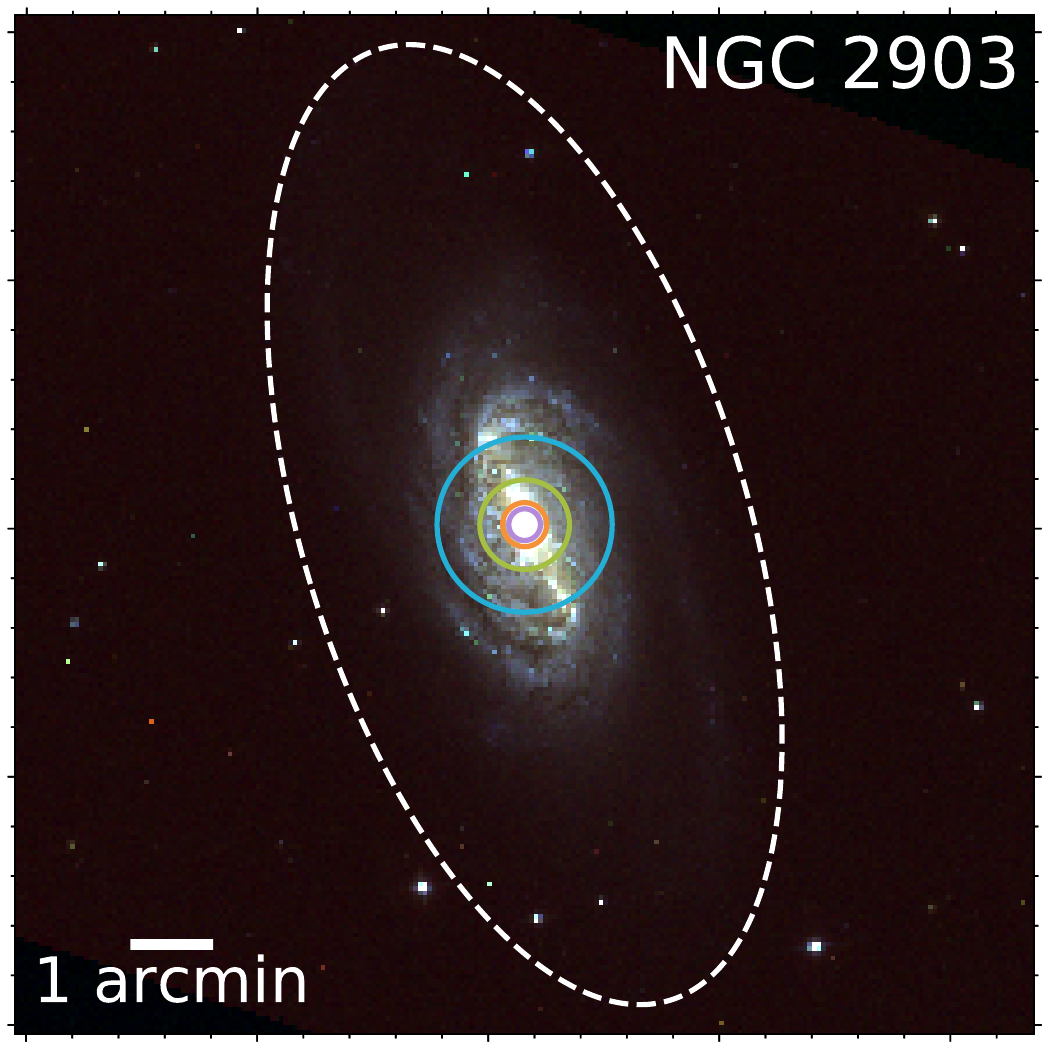} & \hspace{-0.5cm}
    \includegraphics[width=.2\linewidth]{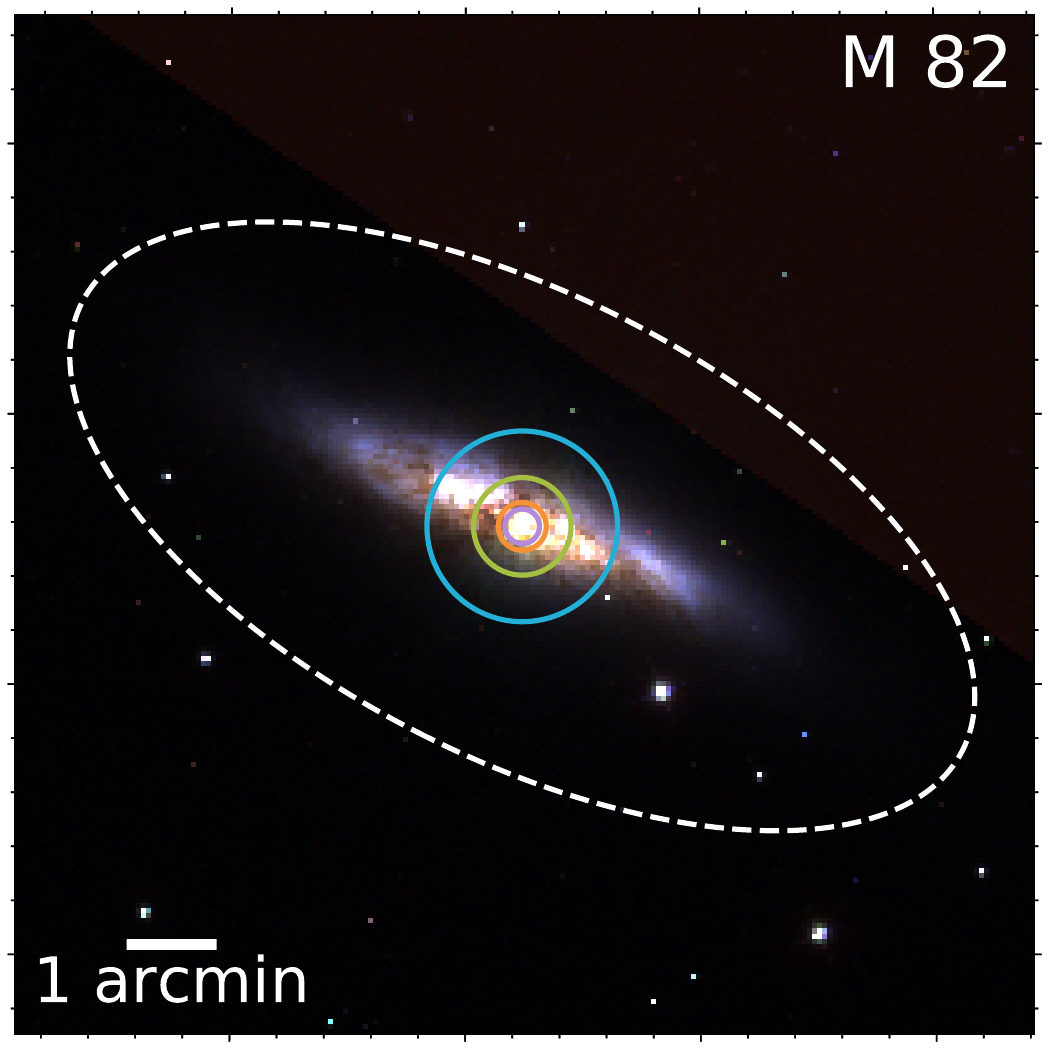} & \hspace{-0.5cm}
    \includegraphics[width=.2\linewidth]{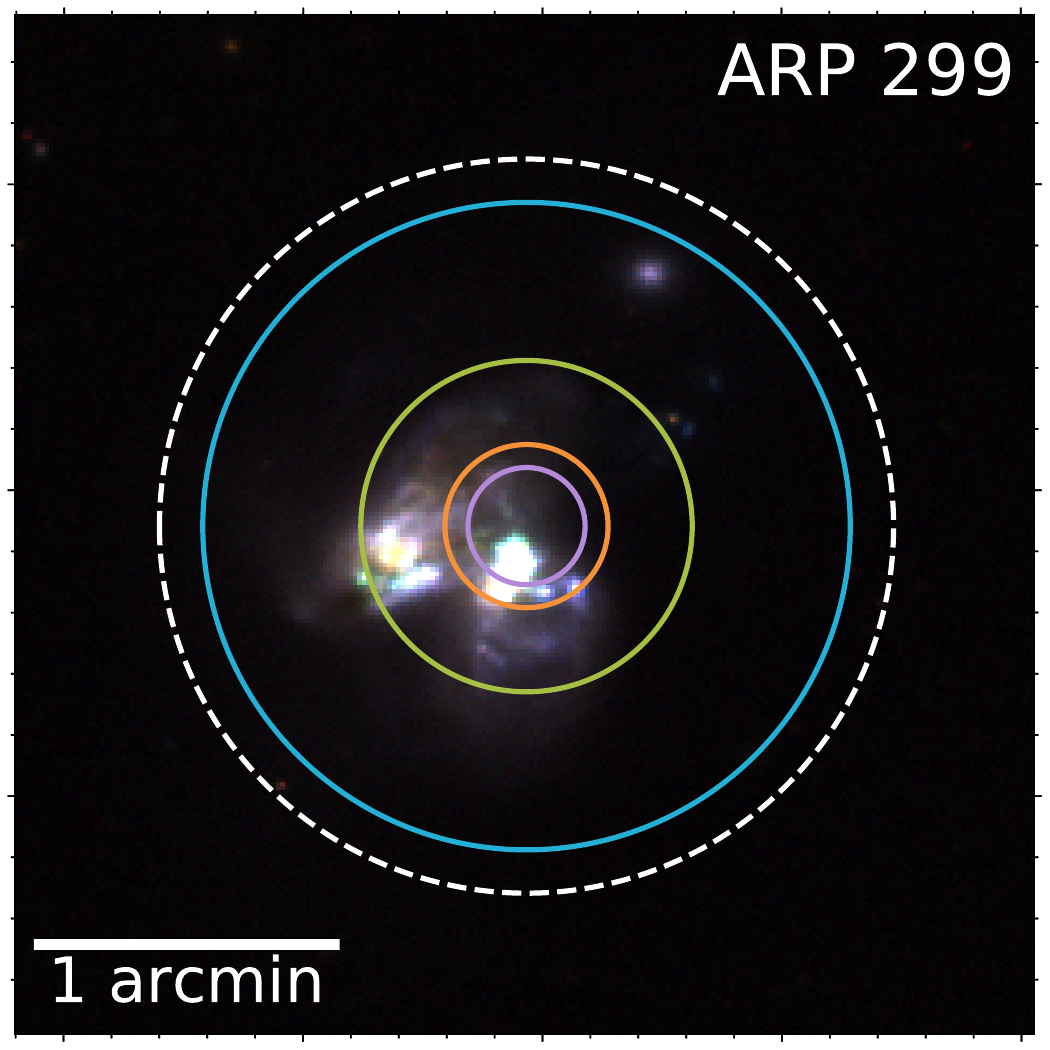} 
  \end{tabular}
  \caption{Optical color images of the sample from the Sloan Digital Sky Survey (SDSS) DR7 (NGC 1068, NGC 2903, M82, and Arp 299) and the 2nd Digitized Sky Survey (DSS2) (NGC 2146). The name of the target is shown in the upper-right corner, and the 1-arcmin scale bar is shown in the lower-left corner. The cyan, green, orange, and purple circles are comparable to the KVN beams at the K, Q, W, and D bands, respectively. Further, the dashed white ellipse is the $D_{25}$ aperture used to measure the integrated emission in multi-frequencies.}\label{fig:optic}
\end{figure*}

This study was focused on only five targets - NGC 1068, NGC 2146, NGC 2903, M82, and Arp 299, which were detected in all four bands to facilitate a more conservative search for AME. The general properties of the five galaxies are presented in Table \ref{tab:sample}. The distance to the galaxies is presented in the units of Mpc, the isophote at the brightness of 25 B$-$mag arcsec$^{-2}$ (D$_{25}$) is presented in the units of arcmin, and the position angle (PA) and inclination are presented in degrees. The optical color images of the sample overlaid with KVN beams and D$_{25}$ are shown in Figure \ref{fig:optic} 

\begin{deluxetable*}{lcccccc}
\tablewidth{0pt}
\tablecaption{The basic properties of the sample galaxies}
\tablehead{Galaxy Name & Distance  & D$_{25}^*$  & PA$^*$ & Inclination$^*$  & Classifications  & Note$^{**}$ \\
 & [Mpc] & [arcmin] & [deg] & [deg] &  & 
}
\startdata 
NGC 1068 & 10.1$^a$ & 6.2 & 70 & 34.7 & AGN$^ 1 $& (R)SA(rs)b; Sy2 \\
NGC 2146 & 18.0$^b$ & 5.4 & 123 & 37.4 & SF$^ 2 $& SB(s)ab pec; HII; LIRG \\
NGC 2903 & 9.2$^c$ & 12.0 & 17 & 67.1 & SF$^ 3 $& SAB(rs)bc I-II HII \\
M82 & 3.5$^d$ & 11.0 & 63 & 76.9 & SF$^ 4 $& I0; Sbrst HII \\
Arp 299 & 54.1$^e$ & 2.4 & ... & ... & SF/AGN$^ 5 $& Interacting system \\
\enddata
\tablecomments{The optical size (D$_{25}^*$), position angle (PA$^*$), inclination, and morphological type have been adopted from the HyperLeda ($^*$\url{https://leda.univ-lyon1.fr}), and the NASA/IPAC Extragalactic Database ($^{**}$\url{https://ned.ipac.caltech.edu}), respectively. The distance to each target has been adopted from the following references: $^a$\citet{Nasonova_2011}; $^b$\citet{Adamo_2012}; $^c$\citet{2016AJ....152...50T}; $^d$\citet{Dalcanton_2009}; $^e$\citet{Gao_2022}. The optical classification has been adopted from the following references: $^1$\citet{Kawakatu_2004}; $^2$\citet{Tabatabaei_2017}; $^3$\citet{Ho_1997}; $^4$\citet{Matsumoto_2001}; $^5$\citet{Gallais_2004}.}
\label{tab:sample}
\end{deluxetable*}

\subsection{KVN spectral data reduction and flux density calibration}
The data were reduced using the GILDAS/CLASS software \citep{2005sf2a.conf..721P, 2013ascl.soft05010G}. For each target, bad scans (e.g., fluctuating baselines under bad weather) were flagged and co-added using sigma weighting of all scans in the unit of measured antenna temperature $(T^*_{\rm A})$, which was corrected for the atmospheric attenuation and the gain variation. The integration times for KVN observations are dependent on the brightness of the target at each frequency, with variations ranging as approximately 1–3, 1–6, 2–5, and 2–6 h in the K-, Q-, W-, and D-bands, respectively, to achieve a sensitivity of approximately 2–12 mK at signal-to-noise ratios of 3 or higher. In addition, the KVN system temperatures also varied across the different frequency bands, ranging as approximately 70–90, 85–140, 150–200, and 140–200 K in the K-, Q-, W-, and D-bands, respectively. The root mean square (RMS) noise levels exhibited variations of 10–25, 20–80, 40–60, and 30–90 mJy in the K-, Q-, W-, and D-bands, respectively. On the final co-added cross-scan spectrum for each target, we fitted a single Gaussian after subtracting the sky background. The Gaussian fitting provided measured antenna temperature, pointing offset, and spectrum width. Consequently, the measured antenna temperature along the RA $(T^*_{\rm A,RA})$ and Dec $(T^*_{\rm A, Dec})$ directions were corrected for the pointing offset by applying the following equations:

\begin{equation}
(T^*_{\rm A,RA})^{'} = T^*_{\rm A,RA}.exp[4 ln2\frac{x^2_{\rm Dec}}{\theta^2_{\rm Dec}}]
\label{eq:ta_ra}
\end{equation}

\begin{equation}
(T^*_{\rm A,Dec})^{'} = T^*_{\rm A,Dec}.exp[4 ln2\frac{x^2_{\rm RA}}{\theta^2_{\rm RA}}]
\label{eq:ta_dec}
\end{equation}
where $(T^*_{\rm A,RA})^{'}$ and $(T^*_{\rm A,Dec})^{'}$ are the corrected antenna temperatures along the RA and Dec directions, respectively, $x_{\rm RA}$ and $x_{\rm Dec}$ are the pointing offsets, respectively, and $\theta_{\rm RA}$ and $\theta_{\rm Dec}$ are the half-power beam width (arcseconds) along the RA and Dec directions, respectively.

The final antenna temperature and its error were estimated by considering the mean of two directions as
\begin{equation}
(T^{*}_{\rm A})^{'} = \frac{(T^{*}_{\rm A,RA})^{'}+(T^{*}_{\rm A,Dec})^{'}}{2}]
\label{eq:ta_aver}
\end{equation}

\begin{equation}
\sigma_{(T^{*}_{\rm A})^{'}} = (T^{*}_{\rm A})^{'}\sqrt{[ \frac{ \sigma^{2}_{\Delta G}}{\Delta G^{2}}+\frac{\sigma^{2}_{T^{*}_{\rm A}}}{(T^{*}_{\rm A})^{2}}]}
\label{eq:err_ta}
\end{equation}
where $\sigma_{(T^*_{\rm A})^{'}}$ is the uncertainty of $(T^*_{\rm A})^{'}$, $\Delta G$ is the antenna gain variation that is estimated by averaging the mean of $(T^*_{\rm A})^{'}$ and $(T^*_{\rm A})^{'}$ of flux calibrators, and $\sigma_{\Delta G}$ is the uncertainty of the gain within the elevation range of 20$^{\circ}$–80$^{\circ}$ from the KVN status report. We utilized well-known flux calibrators such as 3C 286 or planets to determine the antenna gain, which was then used as the conversion factor. This facilitated the conversion of the measured antenna temperature of the source into the intrinsic flux density of that source. The gain factor at a particular frequency is estimated by dividing the flux density of the flux calibrator by the measured antenna temperature of the same flux calibrator. In the final step of our KVN observation, we multiplied the gain factor with the measured antenna temperature of our galaxies to obtain the measured flux density of our observed sample. The KVN flux density of the sample and the error estimate converted into mJy from the monitoring of 3C 286 and planets are presented in Table \ref{tab:KVN_flux}.

\begin{table*}
\centering
\caption{KVN flux density}
\hspace{-1.6cm}
\begin{tabular}{c c r r r r r}
\hline
\multicolumn{1}{c}{\multirow{2}{*}{Band}} & \multicolumn{1}{c}{\multirow{2}{*}{Frequency[GHz]}} & \multicolumn{5}{c}{KVN Flux density [mJy]} \\
\cline{3-7}
\multicolumn{1}{c}{} & \multicolumn{1}{c}{}  & NGC 1068 & NGC 2146 & NGC 2903 & M82 & Arp 299 \\
\hline
K    & 22  & 473$\pm$10 & 114$\pm$3  & 59$\pm$3  & 1355$\pm$13 & 79$\pm$3  \\
Q    & 43  & 218$\pm$18 & 126$\pm$13 & 77$\pm$19 & 890$\pm$41 & 73$\pm$10 \\
W    & 86  & 152$\pm$23 & 163$\pm$23 & 187$\pm$23 & 382$\pm$14 & 91$\pm$25 \\
D    & 129 & 131$\pm$51 & 194$\pm$31 & 69$\pm$20 & 373$\pm$33 & 132$\pm$44 \\ \hline
\end{tabular}
\label{tab:KVN_flux}
\end{table*}

\section{ANCILLARY DATA} \label{sec:anc_data}
To search for the signature of AME, the SED covering a frequency range that is sufficiently broad to include the potential AME bump must be investigated. In this Section, we describe the archival data used to construct SEDs from low-frequency radio to far infrared. In addition, to examine the relationships of AME and other galactic properties including star formation activity, we utilized the archival UV and mid-IR data as described below.

\subsection{Used data for SED fit} \label{subsec:used data sed}
\subsubsection{Low-frequency radio data} \label{subsubsec:radio}
We collected low-frequency radio data from several facilities, including the NRAO Very Large Array (VLA), Westerbork Synthesis Radio Telescope (WSRT), Effelsberg 100-m Radio Telescope, NRAO Green Bank Telescope (GBT), and Algonquin 46-m Radio Telescope (ARO). The list of facilities used is summarized in Table \ref{tab:radio_fac}. The flux densities were measured at multiple frequencies using both interferometric and single-dish observations. To avoid confusion, we mainly considered the flux from the interferometric observations. However, if no synthesis imaging data were available, we made use of the single-dish measurements from the literature. Meanwhile, our targets were very nearby; thus, the contamination was not expected to be serious. 

\begin{table}[h!]
\centering
\caption{Facilities for radio observation}
\hspace{-1.0cm}
\begin{tabular}{lll}
\hline
Telescope	&	Frequency [GHz]	&  Resolution	\\ \hline
VLA	&	1$\sim$2	&	1”$\sim$46”	\\
	&	4$\sim$8	&	0.3”$\sim$12”	\\
	&	8$\sim$12	&	0.2”$\sim$7”	\\
WSRT	&	1.4$\sim$1.7	&	10”$\sim$13”	\\
GBT	&	2.7	&	11’	\\
	&	5	&	6’	\\
ARO	&	6.6	&	2.4’	\\
	&	10.7	&	1.5’	\\
Effelsberg 	&	1.4	&	9.4’	\\
	&	4.85	&	2.4’	\\ \hline
\end{tabular}
\label{tab:radio_fac}
\end{table}

When the synthesis imaging data were available from the literature, including NGC 1068 (VLA-1.4 GHz: \citet{1995ApJ...450..559B}), NGC 2146 (VLA-1.4 GHz: \citet{1996ApJS..103...81C}; WSRT-1.4 and 1.7 GHz: \citet{2007AABraun}), NGC 2903 (VLA-1.4 GHz: \citet{1987ApJS...65..485C}; WSRT-1.5 and 1.7 GHz: \citet{2007AABraun}), M82 (VLA-1.4 GHz: \citet{1987ApJS...65..485C}, VLA-8.3 GHz: \citet{2004ApJ...616..783R}; WSRT-1.7 GHz: \citet{2007AABraun}), and Arp 299 (VLA-1.4 GHz: \citet{1995ApJ...450..559B}), we measured the flux using them. If not available, we used raw data from the archives and processed them using the Common Astronomy Software Applications (CASA) package \citep{2007ASPC..376..127M, 2022PASP..134k4501C} version 6.2.0124. The cases where we reduced the raw data were the VLA C-band observations of NGC 1068 (Project ID: AA281) and M82(Project ID: AS799), and the C-, L-, and X-band observations of Arp 299 (Project ID: AY102, AA216, AA117). 

We calibrated the VLA data following the standard procedures. Subsequently, the UV data were imaged and cleaned using the task {\tt TCLEAN}. To ensure that the flux was not missed, the natural weighting was applied. The resolution of the final images varied as $\sim$ 2"–34". To facilitate a coherent comparison between interferometric and single-dish observations, we convolved the data to match the resolution as described in Section \ref{subsec:photomet}. For a sanity check, we compared the flux of the targets for which both measurements were available. In general, approximately $\sim$ 15$\%$ or less of the flux density was missing in the synthesis imaging data at given frequencies, which barely affected the SED fitting results.

However, in the absence of interferometric observational data, calibrated synthesis image from the literature, or raw data from the archives, we simply adopted the observed flux density from the single-dish measurements reported in the literature, including NGC2146 (Effelsberg-1.5 and 5.0 GHz, \citealt{2017ApJ...836..185T}; ARO-6.6 and 10.7 GHz, \citealt{1973AJ.....78...18M}), NGC 2903 (GBT-4.85 GHz, \citealt{1991ApJS...75....1B}; GBT-5.0 GHz, \citealt{1975AJ.....80..771S}), and M82 (GBT-2.7 and 5.0 GHz, \citealt{1980MNRAS.190..903L}). These single-dish flux densities were measured within the entire system as the total flux density, which is consistent with our study. 

\subsubsection{Sub-millimeter and infrared data} \label{subsubsec:IR}
We used infrared data to investigate the higher frequency part of the potential AME bump in the SED. The data were collected from several facilities as follows: the Multiband Imaging Photometer \citep[MIPS;][]{2004ApJS..154...25R} on Spitzer, the Fourier Transform Spectrometer of the Spectral and Photometric Imaging Receiver \citep[SPIRE;][]{2010A&A...518L...3G}, the Photodetector Array Camera and Spectrometer \citep[PACS;][]{2010A&A...518L...2P} on the Herschel Space Observatory, the Infrared Astronomical Satellite \citep[IRAS;][]{1984ApJ...278L...1N}, and the Submillimetre Common-User Bolometer Array 2 \citep[SCUBA-2;][]{2013MNRAS.430.2513H} on JCMT. The sources of the data and the references are summarized by the target in Table \ref{tab:IRdata}.

\renewcommand{\arraystretch}{1.2} 
\begin{table*}
\renewcommand{\arraystretch}{1.2}
\tabletypesize{\scriptsize}
\caption{Sub-millimeter and infrared data}
\label{tab:IRdata}
\hspace{-1.5cm}
\begin{tabular}{l l l c c c c c } \hline \hline
Facility & Wavelength [$\mu$m] & Tracer & NGC 1068 & NGC 2146 & NGC 2903 & M82 & Arp 299\\ \hline
JCMT/SCUBA2	    & 450, 850       & thermal dust   & a	&	…	&	b	&	c	&	d	\\
Herschel/SPIRE	& 250, 350, 500  & thermal dust   &	e	&	f	&	…	&	e	&	g	\\
Herschel/PACS	& 70, 100, 160   & thermal dust   &	e	&	f	&	…	&	e	&	g	\\
Spitzer/MIPS    & 70, 160        & thermal dust   &	m	&	…	&	j	&	…	&	…	\\
Spitzer/MIPS    & 24             & hot small dust & m	&	…	&	j	&	…	&	…	\\
Spitzer/IRAC	& 8.0            & PAHs           &	h	&	i	&	j	&	k	&	l	\\
Spitzer/IRAC	& 3.6            & old stars      &	h	&	i	&	j	&	k	&	l	\\
IRAS/IRIS/HIRES & 60, 100        & thermal dust   &	n	&	o   &	n	&	n	&	k	\\
IRAS/IRIS/HIRES & 25             & hot small dust &	n	&	o   &	n	&	n	&	k	\\
IRAS/IRIS/HIRES & 12             & PAHs           &	n	&	o   &	n	&	n	&	k	\\ \hline
\end{tabular}
\tablecomments{
$^{a}$SCUBA-2/M97BC27;
$^{b}$SCUBA-2/M03AC13;
$^{c}$SCUBA-2/MJLSN06,MJLSN07;
$^{d}$SCUBA-2/M97BI27;
$^{e}$\citet{2012MNRAS.419.1833B};
$^{f}$\citet{2011PASP..123.1347K};
$^{g}$\citet{2013ApJ...768...90L};
$^{h}$\citet{2014ApJS..212...18B};
$^{i}$\citet{2004sptz.prop...59R};
$^{j}$\citet{2009ApJ...703..517D};
$^{k}$\citet{2003PASP..115..928K};
$^{l}$\citet{2004sptz.prop...32F};
$^{m}$\citet{2012MNRAS.423..197B};
$^{n}$\citet{Miville-Deschenes_2005};
$^{o}$\citet{2004AJ....127.3235S}}
\end{table*}

\subsection{Non-used data for SED fit}
To trace the dust properties, we collected data from the InfraRed Array Camera \citep[IRAC;][]{Fazio_2004} on Spitzer. The sub-millimeter and infrared data are summarized in Table \ref{tab:IRdata}. In addition, we used the far-ultraviolet (FUV, $\lambda\approx$ 1528 \r{A}) and near-ultraviolet photometric data (NUV, $\lambda\approx$ 2271 \r{A}) from the Galaxy Evolution Explorer \citep[GALEX;][]{2005ApJ...619L...1M} archive (GR6) Nearby Galaxy Survey \citep[NGS;][]{2003AAS...203.9112B}. Combined with the IR data, these UV data were used to probe the star formation activity of the sample. 

\subsection{The photometry} \label{subsec:photomet}
All the data was obtained with the sky brightness subtracted, and the instrumental noise was incorporated in the uncertainty of the targeted area flux. The synthesis images in low radio frequencies were convolved to match the KVN K-band resolution ($\approx$ 130”) using a Gaussian kernel. In the case of sub-millimeter and IR data, the Spitzer IRAC/MIPS and Herschel PACS data were first convolved to the resolution of the Herschel-SPIRE 500 $\mu$m data using the custom kernels \citep{2011PASP..123.1218A} to enhance consistency among the datasets from various facilities. Consequently, the images of the same target were cropped to a common size of two times the optical $D_{25}$ size ($\approx$ $2D_{25}$), and the Gaussian kernel was used to convolve the data to match the 130” beam of KVN K-band data. In this process for all smoothed images, the pixel sizes were chosen by the CASA command, which is not identical but comparable, and corresponded to $\sim$ 3 - 5 elements across the KVN K-band beam. One exception was the IRAS 100 $\mu$m data that were comparable to or larger than 130” resolution. Using this, we measured the flux using the original images without convolution. Finally, the global flux density was measured within the aperture with the size of $D_{25}$ at all wavelengths for each target.

\section{SED Modeling} \label{sec:sedfit}
In addition to the AME, there are four emission components contributing to the observed radio SED: synchrotron emission, free-free emission, thermal dust emission, and CMB anisotropy component. Consequently, the total flux density is expressed as
\begin{equation}
S_{\rm mod}= S_{\rm syn}+S_{\rm ff}+S_{\rm thd}+S_{\rm CMB}+S_{\rm AME},
\end{equation}
Where $S_{\rm syn}$, $S_{\rm ff}$, $S_{\rm thd}$, $S_{\rm CMB}$, and $S_{\rm AME}$ are the flux densities of the synchrotron emission, free-free emission, thermal dust emission, CMB emission, and AME, respectively. We describe the proposed modeling of these components in the subsequent sections.

\subsection{Synchrotron emission} \label{subsec:syn}
None of our targets exhibited a turnover at low frequencies of our radio SED. Thus, we assumed only an optically thin synchrotron component, whose flux density can be described by the following power law \citep[e.g.,][]{2016era..book.....C}, 
\begin{equation}
S_{\rm syn}(\nu) = A_{\rm syn}\times (\frac{\nu}{1.0\ \rm GHz})^{\alpha_{\rm syn}}
\end{equation}
where $A_{\rm syn}$ is the amplitude of the synchrotron component at 1.0 GHz and $\alpha_{\rm syn}$ is the spectral index of the synchrotron emission. Both $A_{\rm syn}$ and $\alpha_{\rm syn}$ were considered as free parameters in our radio SED data fitting.

\subsection{Free-free emission} \label{subsec:ff}
The free-free thermal emission in the radio regime is described by the Rayleigh-Jeans limit of the Planck function as follows: 
\begin{equation}
S_{\rm ff}(\nu) = \frac{2k_{\rm B}\nu^2}{c^2}\Omega_{\rm b}T_{\rm ff}(\nu)
\end{equation}
where $k_{\rm B}$ is the Boltzmann constant and $\Omega_{\rm b}$ is the solid angle of the source (considered as $D_{25}$ of each target). Further, $T_{\rm ff}$ is the free-free brightness temperature following \citet{2011piim.book.....D}'s model, that is, 

\begin{equation}
T_{\rm ff}(\nu) = T_{\rm e}\times (1-exp[-\tau_{\rm ff}(\nu)])
\end{equation}
where $T_{\rm e}$ is the electron temperature (set as 8000 K, that is, the typical value of the Milky Way). Further, $\tau_{\rm ff}(\nu)$ is the free-free optical depth and is obtained as

\begin{equation}
\tau_{\rm ff}(\nu) = 5.468\times10^{-2}\rm EM\times T^{-\frac{3}{2}}_{\rm e}(\frac{\nu}{\rm GHz})^{-2}g_{\rm ff}(\nu)
\end{equation}
where $EM = \int n^2_{\rm e} dl$ is the emission measure along the line of sight of an HII region of depth $l$. When a single $T_{\rm e}$ was assumed as here, $\tau_{\rm ff}(\nu)$ varied with the electron density $n_{\rm e}$, and a dimensionless free-free Gaunt factor \citep{2011piim.book.....D}, $g_{\rm ff}(\nu)$ as follows,

\begin{equation} \label{eq:gff}
exp[g_{\rm ff}(\nu)] = exp\lbrace 5.960-  \frac{\sqrt{3}}{\pi}\times ln[\frac{\nu}{\rm GHz}(\frac{T_{\rm e}}{10^4\ \rm K})^{-\frac{3}{2}}]\rbrace + e.
\end{equation}
Thus, EM was the only free parameter left in the fitting procedure of the free-free emission component.

\subsection{Thermal dust emission} \label{subsec:thd}
We adopted a modified blackbody model to fit the optically thin thermal dust component as follows:
\begin{equation}
   	  S_{\rm thd}(\nu) = \frac{2h\nu^3}{c^2}(\frac{\nu}{353\ \rm GHz})^{\beta}[\frac{1}{e^{h\nu/k_{\rm B}T_{\rm d}}-1}]\tau_{353}\Omega_{\rm b}
\end{equation}	
where $\beta$ is the emissivity index, $\tau_{353}$ is the optical depth at 353 GHz, and $T_{\rm d}$ is the dust equilibrium temperature. In (sub-)mm-cm wavelengths, the cold dust component, $T_{\rm d}$ $<$ 30 K, is expected to be predominant \citep{10.1111/j.1365-2966.2012.21667.x} whereas the warm dust component of $T_{\rm d}$ $>$ 30 K makes minimal contributions to the SED of the frequency range probed herein. Therefore, we ignored the warm dust component in this study.

\subsection{CMB emission} \label{subsec:cmb}
The CMB component was also included in our SED analysis. The flux density from the CMB component is estimated by the following equation \citep{fernandeztorreiro2023quijote},
\begin{equation}
   	  S_{\rm CMB}(\nu) = \frac{2k_{\rm B}\nu^2}{c^2} \frac{(h\nu/k_{\rm B}T_{\rm CMB})^2 e^{h\nu/k_{\rm B}T_{\rm CMB}}}{(e^{h\nu/k_{\rm B}T_{\rm CMB}}-1)^2}\Delta T_{\rm CMB}\Omega_{\rm b}
\end{equation}	
where $\Delta T_{\rm CMB}$ is the differential temperature of CMB perturbation and $T_{\rm CMB}$ is the CMB temperature, which is a fixed value of 2.72548 K \citep{2009ApJ...707..916F}.

\subsection{Anomalous microwave emission} \label{subsec:ame}
We adopted the parametric model, as in \citet{Cepeda-Arroita_2021} and \citet{fernandeztorreiro2023quijote}, to describe the AME component in the radio SED, which was first introduced by \citet{Stevenson_2014}. This symmetrical distribution introduces only three free parameters to avoid the challenges associated with high dimensionality and degeneracies encountered in the SpDust (v2) code model \citep{AliHaimoud_2009, Silsbee_2011}; thus, it is more directly applicable to the observational data. The AME component is described as follows,

\begin{equation}
S_{\rm AME}(\nu) = A_{\rm AME}\times exp(-\frac{1}{2}[\frac{\ln(\nu/\nu_{\rm peak})}{W_{\rm AME}}]^2)
\end{equation}	
where $A_{\rm AME}$ is the peak amplitude, $\nu_{\rm peak}$ is the peak frequency, and $W_{\rm AME}$ is the width of the spinning dust spectrum. The fitting process described in \ref{subsec:fitting} returns $A_{\rm AME}$, $\nu_{\rm peak}$, and $W_{\rm AME}$, and the presence of AME is determined by the significance of $S_{\rm AME}$. 

\subsection{Fitting process and validation} \label{subsec:fitting}
The observed SED data ($S_{\rm obs}$) were fitted with the SED model ($S_{\rm mod}$) using the python package {\tt EMCEE} ensemble sampler \citep{2013PASP..125..306F}. The Markov chain Monte Carlo (MCMC) algorithm was used with the maximum likelihood estimation to determine the parameters that best matched the observations statistically. The SED was fitted in the 10- and 7-dimensional parameter space for the models with and without AME, respectively. We used 300 walkers and then executed 10,000 steps of MCMC, resulting in well converged results. We neglected 5,000 steps at the beginning to remove the burn-in phase from the chain. The boundary of free parameters was adopted from \citet{fernandeztorreiro2023quijote} except $\nu_{\rm peak}$, where $\nu_{\rm peak}$ was extended to vary as 10–90 GHz. The boundaries used were
0 $\leq$ $A_{\rm syn}$ [Jy] $\leq$ $+inf$,
-2 $\leq$ $\alpha_{\rm syn}$ $\leq$ 1,
1 $\leq$ $\rm EM$ [$\rm cm^{-6}pc$] $\leq$ $+inf$,
10 $\leq$ $T_{\rm d}$ [K] $\leq$ 40,
0 $\leq$ $\beta$ $\leq$ 3,
-7 $\leq$ $\tau_{353}$ $\leq$ -1,
-600 $\leq$ $\Delta T_{\rm CMB}$ [$\mu \rm K$] $\leq$ 600,
0 $\leq$ $A_{\rm AME}$ [Jy] $\leq$ $+inf$,
10 $\leq$ $\nu_{\rm peak}$ [GHz] $\leq$ 90, and
0.2 $\leq$ $W_{\rm AME}$ $\leq$ 1.
To obtain initial guesses for the MCMC fit, we first used the data of the frequency $\nu$ $\lesssim$ 10 GHz to fit synchrotron and free-free components while the data of $\nu$ $\gtrsim$ 353 GHz were used to fit the thermal dust emission. Thereafter, the best-fit parameters were used as the inputs for the full radio-IR SED fitting based on the prescriptions described in Sections \ref{subsec:syn}-\ref{subsec:ame} with the constraint we have mentioned before. The statistical significance of the SED fitting with and without the AME component is presented in Table \ref{tab:parameter_conv}. The chi-square $\chi^2$ for each target over the fitted wavelength range has been estimated as follows,

\begin{table*}[h!]
\centering
\caption{Best-fitted parameters}
\hspace{-1.6cm}
\begin{tabular}{lrrrrr}
\hline \hline
\multicolumn{6}{c}{A model including spinning dust}  \\
\hline
Parameter	&	NGC 2903	&	NGC 2146	&	 M82 &      NGC 1068	&	Arp 299	  \\ \hline 
${A_{\rm syn}\ [{\rm Jy}]}$ 	 & $	0.71_{-0.01}^{+0.01}	$ & $	 1.48_{-0.05}^{+0.05}	$ & $	 12.96_{-0.08}^{+0.08}	$ & $	8.54_{-0.27}^{+0.29}	$ & $	0.23_{-0.10}^{+0.11}	$ \\
${\alpha_{\rm syn}}$ 	 & $	-1.01_{-0.08}^{+0.07}	$ & $	 -0.85_{-0.03}^{+0.02}	$ & $	 -0.87_{-0.00}^{+0.00}	$ & $	-0.94_{-0.02}^{+0.02}	$ & $	-1.12_{-0.61}^{+0.57}	$ \\
${{\rm EM}\ [\rm cm^{-6}pc]}$ 	 & $	83_{-31}^{+26}	$ & $	 42_{-32}^{+63}	$ & $	 12_{-9}^{+20}	$ & $	65_{-49}^{+97}	$ & $	2700_{-1163}^{+649}	$ \\
${T_{\rm d}\ [\rm K]}$ 	 & $	23.6_{-1.0}^{+1.1}	$ & $	 30.1_{-1.5}^{+1.6}	$ & $	 28.0_{-1.0}^{+1.0}	$ & $	27.4_{-1.0}^{+1.1}	$ & $	33.8_{-1.5}^{+1.5}	$ \\
${\beta}$ 	 & $	2.27_{-0.15}^{+0.14}	$ & $	 1.84_{-0.13}^{+0.13}	$ & $	 2.38_{-0.08}^{+0.08}	$ & $	1.82_{-0.12}^{+0.12}	$ & $	1.98_{-0.10}^{+0.11}	$ \\
${log_{10}(\tau_{353})}$ 	 & $	-5.52_{-0.01}^{+0.01}	$ & $	 -5.03_{-0.02}^{+0.02}	$ & $	 -4.82_{-0.02}^{+0.02}	$ & $	-4.76_{-0.03}^{+0.03}	$ & $	-5.09_{-0.02}^{+0.02}	$ \\
${\Delta T_{\rm CMB}\ [\mu \rm K]}$ 	 & $	-8_{-47}^{+30}	$ & $	 172_{-117}^{+88}	$ & $	 112_{-18}^{+18}	$ & $	-98_{-88}^{+68}	$ & $	-100_{-300}^{+300}	$ \\
${A_{\rm AME}\ [\rm Jy]}$ 	 & $	0.18_{-0.04}^{+0.05}	$ & $	 0.09_{-0.04}^{+0.04}	$ & $	 0.84_{-0.16}^{+0.38}	$ & $	0.03_{-0.02}^{+0.05}	$ & $	0.03_{-0.02}^{+0.03}	$ \\
${\nu_{\rm AME}\ [\rm GHz]}$ 	 & $	76.7_{-7.9}^{+8.6}	$ & $	 70.8_{-10.1}^{+13.6}	$ & $	 30.0_{-0.6}^{+0.4}	$ & $	72.7_{-55.0}^{+13.4}	$ & $	21.6_{-8.3}^{+56.7}	$ \\
${W_{\rm AME}}$ 	 & $	0.32_{-0.08}^{+0.11}	$ & $	 0.39_{-0.12}^{+0.11}	$ & $	 0.28_{-0.05}^{+0.05}	$ & $	0.37_{-0.13}^{+0.36}	$ & $	0.47_{-0.22}^{+0.38}	$ \\ \hline
$\chi^2$ 	 & $	61.7	$ & $	35.6 $ & $	1301.0	$ & $	50.3	$ & $	35.0	$ \\
$\chi_{n}^2$ 	 & $	15.4	$ & $	4.5	$ & $	185.9 $ & $	16.8	$ & $	5.8	$ \\
$AIC$ 	 & $	222.5	$ & $	288.0 $ & $	1559.5	$ & $	266.1	$ & $	256.7	$ \\ \hline
$S/N_{\rm AME}$ 	 & $	4.5	$ & $	2.3	$ & $	2.2	$ & $	0.6	$ & $	1.0	$ \\ \hline
\hline \hline
\multicolumn{6}{c}{A model without spinning dust}  \\ \hline 
Parameter	&	NGC 2903	&	NGC 2146	&	 M82 &      NGC 1068	&	Arp 299	  \\ \hline 
${A_{\rm syn}\ [\rm Jy]}$ 	 & $	0.71_{-0.01}^{+0.01}	$ & $	 1.48_{-0.05}^{+0.05}	$ & $	 12.14_{-0.09}^{+0.09}	$ & $	8.51_{-0.27}^{+0.28}	$ & $	0.20_{-0.09}^{+0.10}	$ \\
${\alpha_{\rm syn}}$ 	 & $	-1.03_{-0.07}^{+0.07}	$ & $	 -0.86_{-0.03}^{+0.02}	$ & $	 -1.00_{-0.01}^{+0.01}	$ & $	-0.94_{-0.02}^{+0.02}	$ & $	-0.98_{-0.70}^{+0.52}	$ \\
${{\rm EM}\ [\rm cm^{-6}pc]}$ 	 & $	83_{-27}^{+23}	$ & $	 53_{-39}^{+76}	$ & $	 2327_{-95}^{+93}	$ & $	66_{-50}^{+100}	$ & $	3340_{-1326}^{+282}	$ \\
${T_{\rm d}\ [\rm K]}$ 	 & $	22.3_{-0.8}^{+0.8}	$ & $	 29.9_{-1.4}^{+1.6}	$ & $	 28.9_{-1.0}^{+1.1}	$ & $	27.3_{-1.0}^{+1.1}	$ & $	33.9_{-1.4}^{+1.5}	$ \\
${\beta}$ 	 & $	2.48_{-0.11}^{+0.12}	$ & $	 1.87_{-0.13}^{+0.13}	$ & $	 2.29_{-0.08}^{+0.08}	$ & $	1.83_{-0.12}^{+0.12}	$ & $	1.97_{-0.09}^{+0.09}	$ \\
${log_{10}(\tau_{353})}$ 	 & $	-5.53_{-0.01}^{+0.01}	$ & $	 -5.04_{-0.03}^{+0.02}	$ & $	 -4.81_{-0.02}^{+0.02}	$ & $	-4.76_{-0.03}^{+0.03}	$ & $	-5.09_{-0.02}^{+0.02}	$ \\
${\Delta T_{\rm CMB}\ [\mu \rm K]}$ 	 & $	57_{-14}^{+14}	$ & $	 315_{-48}^{+49}	$ & $	 -384_{-27}^{+27}	$ & $	-60.1_{-51.5}^{+50.7}	$ & $	-242_{-193}^{+219}	$ \\ \hline
$\chi^2$ 	 & $	92.0	$ & $	46.0	$ & $	1848.0	$ & $	49.4	$ & $	31.8	$ \\
$\chi_{n}^2$ 	 & $	11.5	$ & $	3.8	$ & $	168.0	$ & $	7.1	$ & $	3.2	$ \\
$AIC$ 	 & $	248.9	$ & $	294.4	$ & $	2102.4	$ & $	261.2	$ & $	249.5	$ \\ \hline
$LR-statistic$ 	 & $	30.4	$ & $	10.4	$ & $	546.9	$ & $	-0.9	$ & $	-3.2	$ \\
p-value  	 & 	 $\ll 0.05$ 	&	$0.02$	&	 $\ll 0.05 $ 	&	 $1.00$ 	&	 $1.00$  \\ \hline	
Best-Model &  SpD  &  SpD  &  SpD  &  No-SpD & 	No-SpD		 \\ \hline
\end{tabular}
\label{tab:parameter_conv}
\end{table*}

\begin{equation}
\chi^2 = \Sigma{\frac{1}{\sigma_i^2}[y_i - y(x_i)]^2}
\end{equation}
where $y_i$ and $y(x_i)$ are the observed and fitted flux densities, respectively, and $\sigma_i$ is the uncertainty of the observed flux density. The reduced $\chi^2$ is defined as $\chi^2$ divided by the degrees of freedom $n$ (i.e., $\chi^2_{n}$ = $\chi^2/n$).

We also performed the likelihood ratio test to verify the statistical significance of the SED fitting with and without AME. We estimated the $LR$ statistical number using
\begin{equation} \label{eq:lr}
LR = -2\ln\frac{\mathcal{L}_0}{\mathcal{L}} 
\end{equation}
where $\mathcal{L}$ and $\mathcal{L}_0$ are the log-likelihood of the best-fit model with and without the AME component, respectively. In addition, the Akaike Information Criterion ($AIC$) test was conducted. The $AIC$ is defined as
\begin{equation} \label{eq:aic}
AIC = 2K - 2\ln \mathcal{L}
\end{equation}
where $K$ is the number of independent variables and $\mathcal{L}$ is the same as in Eq. \ref{eq:lr}, that is, the log-likelihood. The smaller the $AIC$, the better the SED fit. 

\begin{figure*}
\centering
    \includegraphics[width=.46\linewidth]{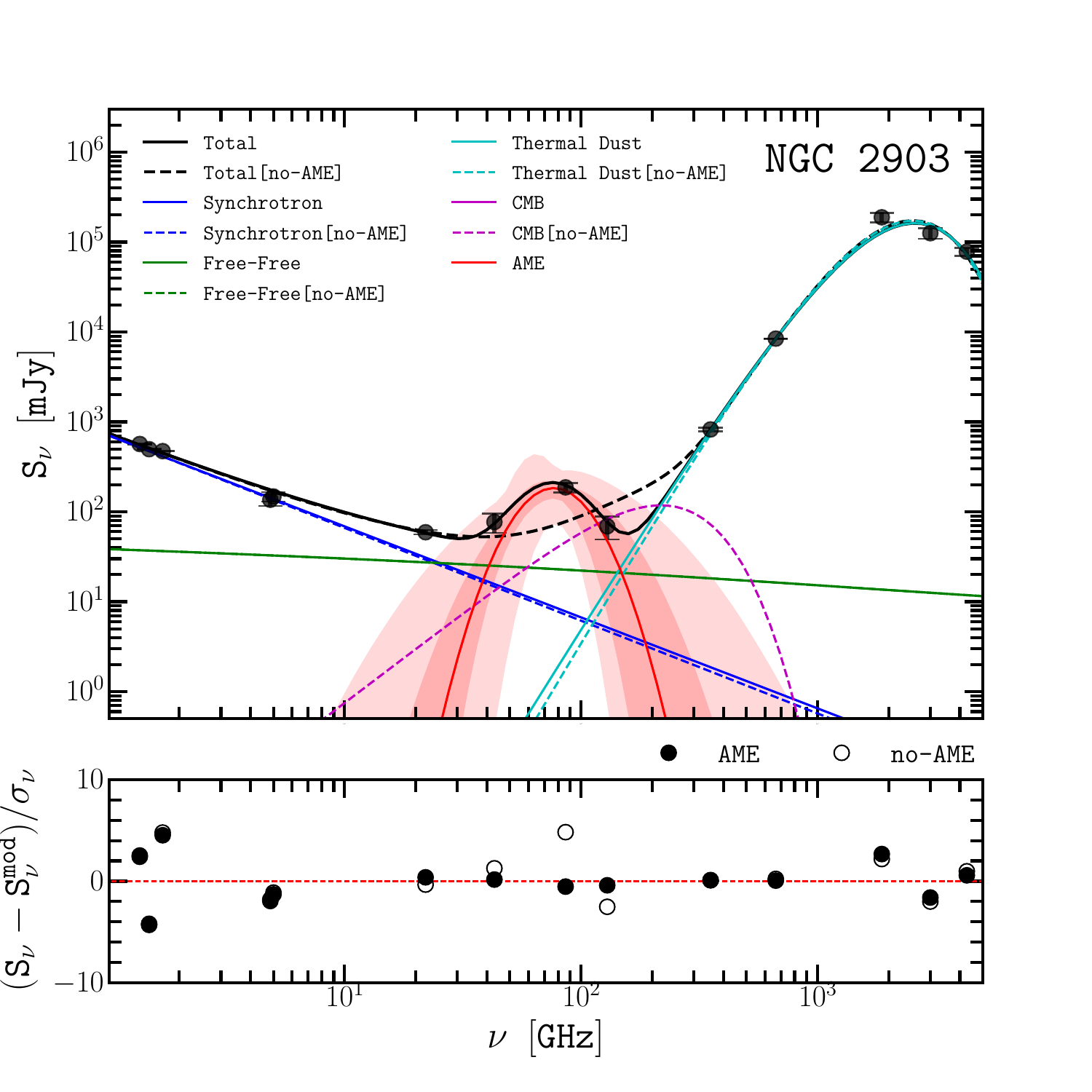}
    \includegraphics[width=.46\linewidth]{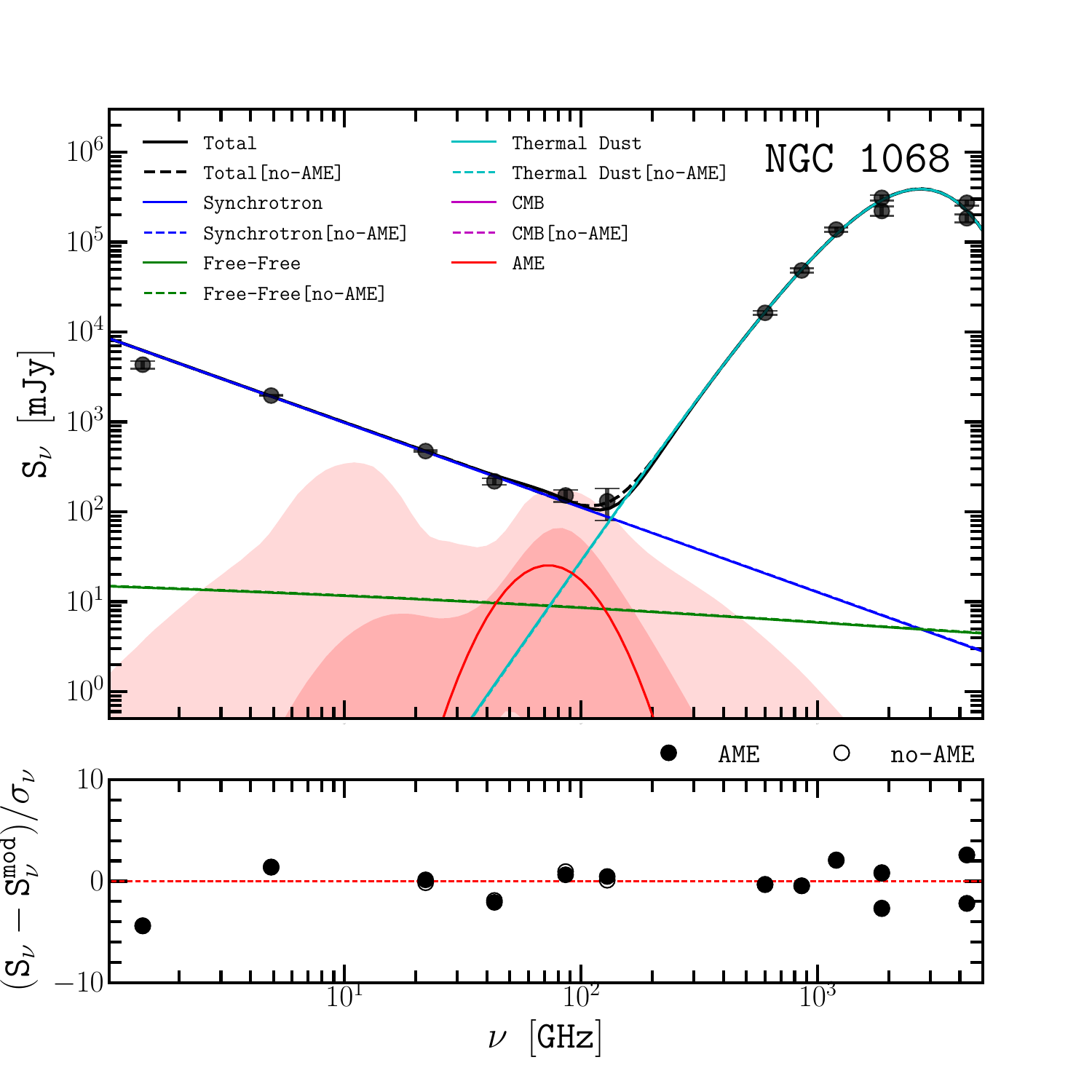}
    \\ \vspace{-0.3cm}
    \includegraphics[width=.46\linewidth]{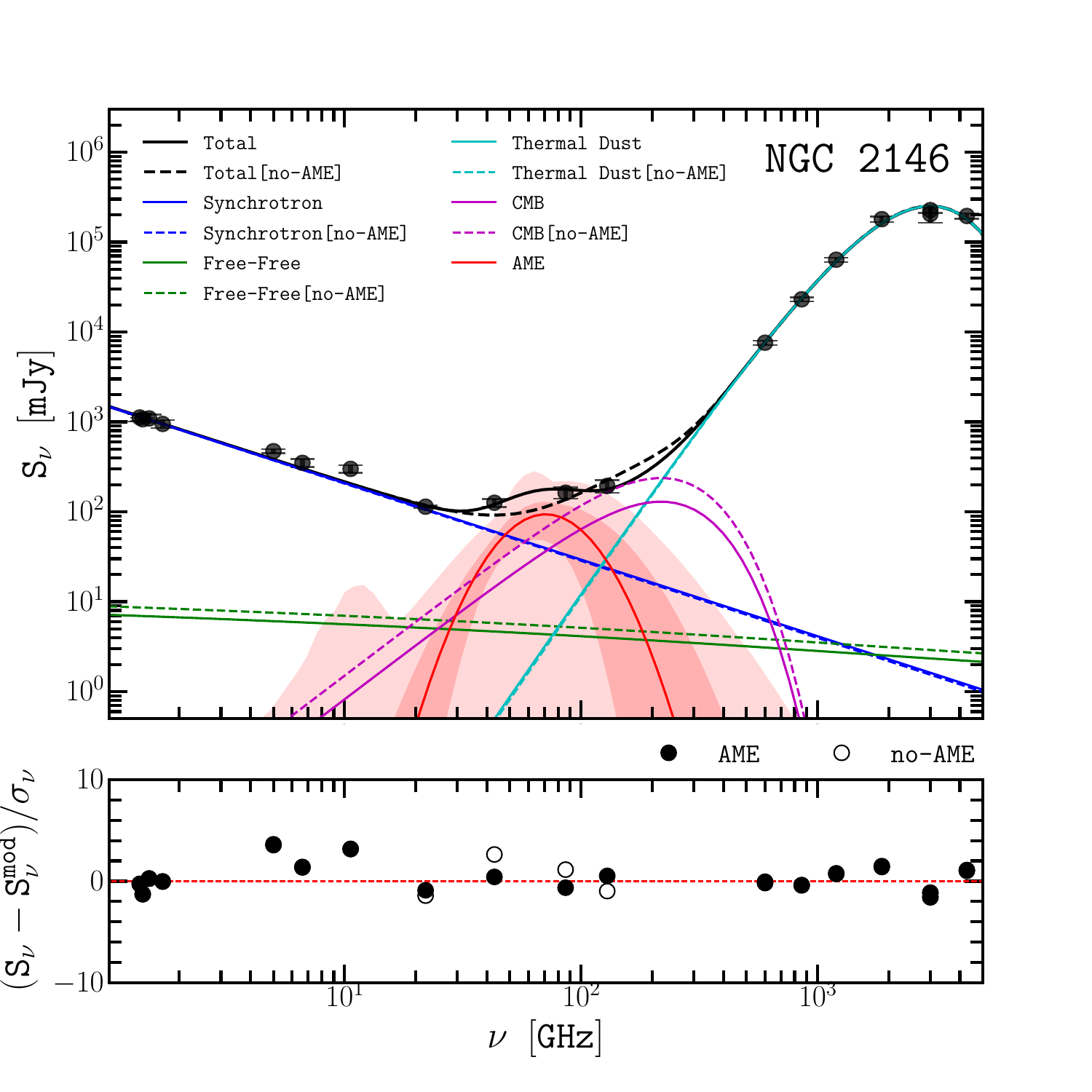}
    \includegraphics[width=.46\linewidth]{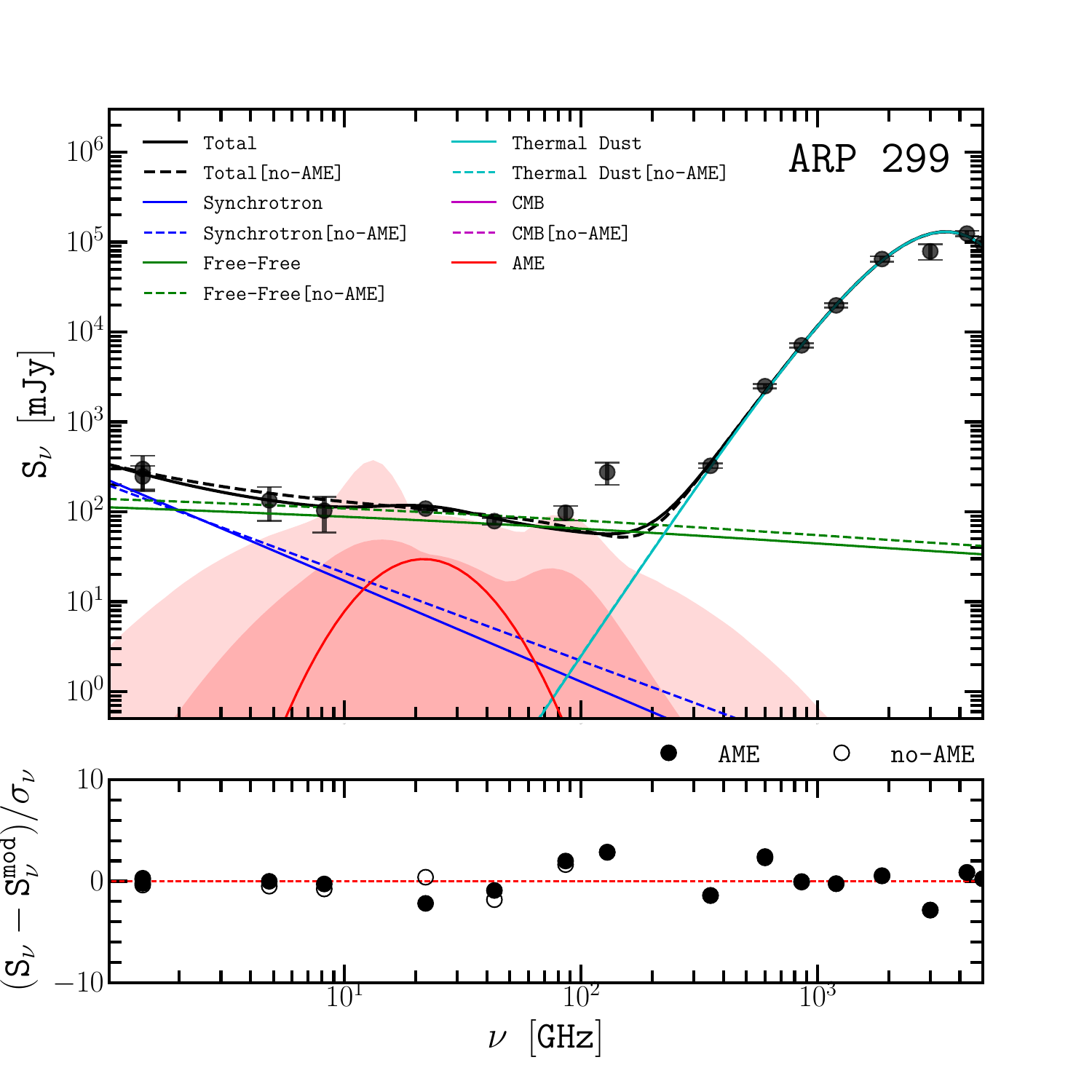}
    \\ \vspace{-0.3cm}
    \includegraphics[width=.46\textwidth]{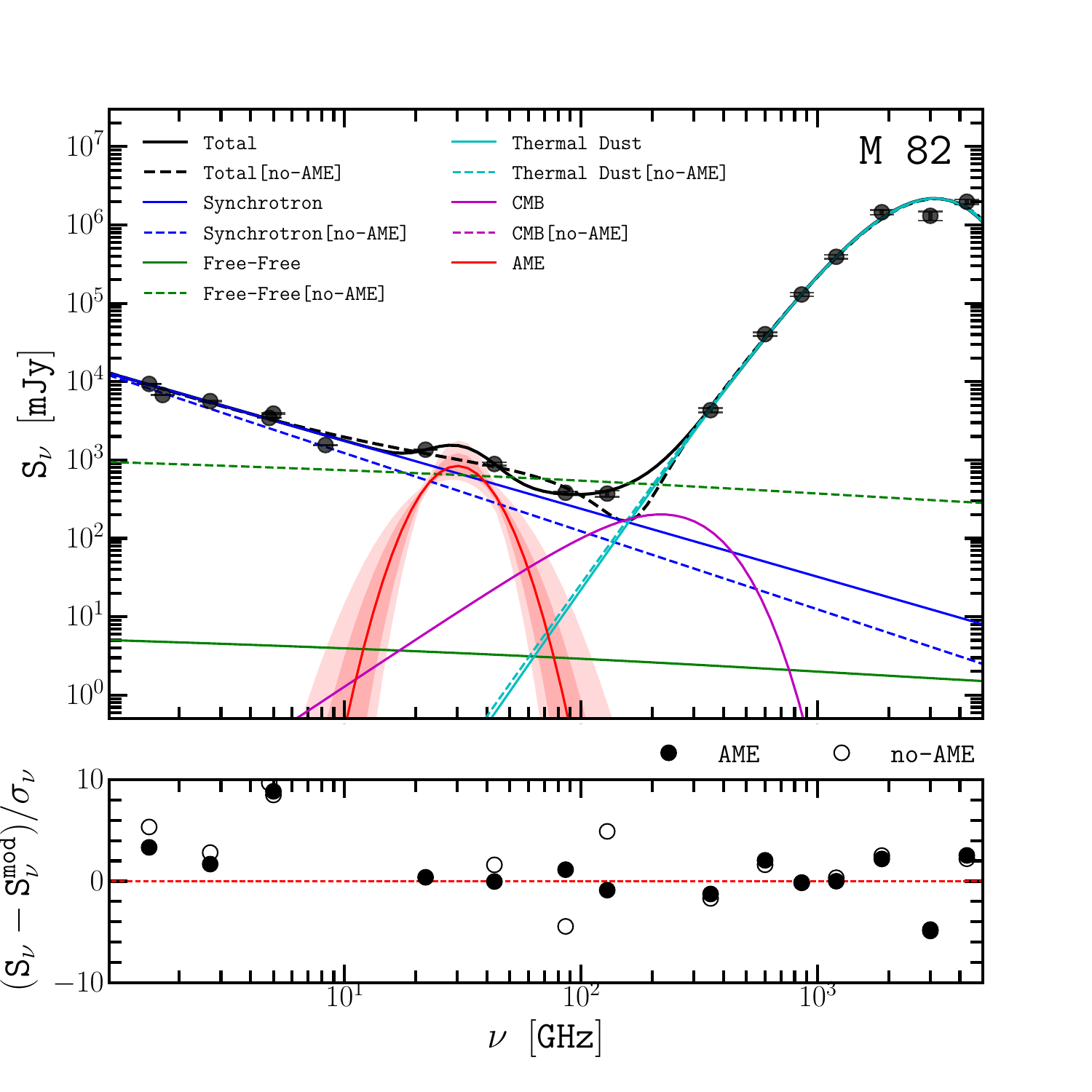} 
    \begin{minipage}{.46\linewidth}
\vspace{-9cm} \caption{Top: global SEDs with the fitting results of the galaxy samples using the flux density extracted from the convolved data. The observed data (solid dots), fitted SED with AME (solid line), and fitted SED without AME (dashed line). The synchrotron, free-free, thermal dust, and  CMB anisotropy components are shown by blue, green, cyan, and magenta solid lines (using a model including spinning dust) and dashed lines (using a model without spinning dust), respectively. The red solid line is the AME component assuming the spinning dust model with a shaded area of its one and three-sigma uncertainties. Bottom: signal-to-noise ratios of the residuals for the with (solid black circles) and without AME (open circles) models. The targets likely and unlikely to contain AME are shown on the left and on the right columns, respectively.}\label{fig:sed}
\end{minipage}
\end{figure*}

The statistical tests suggested that among the sample of five, NGC 2146, NGC 2903, and M82 favored the model with the AME component, which is supported by the $AIC$ and the $LR$ statistical tests with p-values $<$ 0.05. Although the $\chi^2_{n}$ values deny the favor of the model with AME, as evident, the AME-included model described the observed SED better than the model without the AME component. This was particularly observed in the essential frequencies for the AME component, which is supported by smaller $\chi^2$ values in the AME-included model cases. In addition, results based on the convolved data are provided; however, the data of the original resolutions yielded a consistent result, favoring the model with AME in the same three targets. Although there are three galaxies for which the presence of AME was suggestive based on statistical tests, the AME in NGC 2146 and M82 were classified as marginal detections owing to the AME component being embedded in total SED and having a small signal-to-noise ratio, $S/N_{\rm AME}\approx2.0$. Figure \ref{fig:sed} shows the SED fits based on the convolved data. As evident, the emission at low radio frequencies was dominated by the synchrotron component and the dominance decreased with increase in the frequencies. Simultaneously, the free-free component dominated within a narrow range of a few GHz to a few tens of GHz. However, at much higher frequencies, the contributions from the CMB anisotropy and thermal dust emission components were more prominent. Between a few hundred to a few thousand of GHz, thermal dust emission emerged as the most dominant component. 
Notably, AME played a crucial role within the frequency range of 10-100 GHz, particularly in the case of the AME-detected galaxy, NGC 2903.

\section{AME properties}  \label{sec:result}
\subsection{The signature of AME}\label{subsec:sig_ame}
\begin{figure}
	\centering 
	\includegraphics[width=0.8\linewidth]{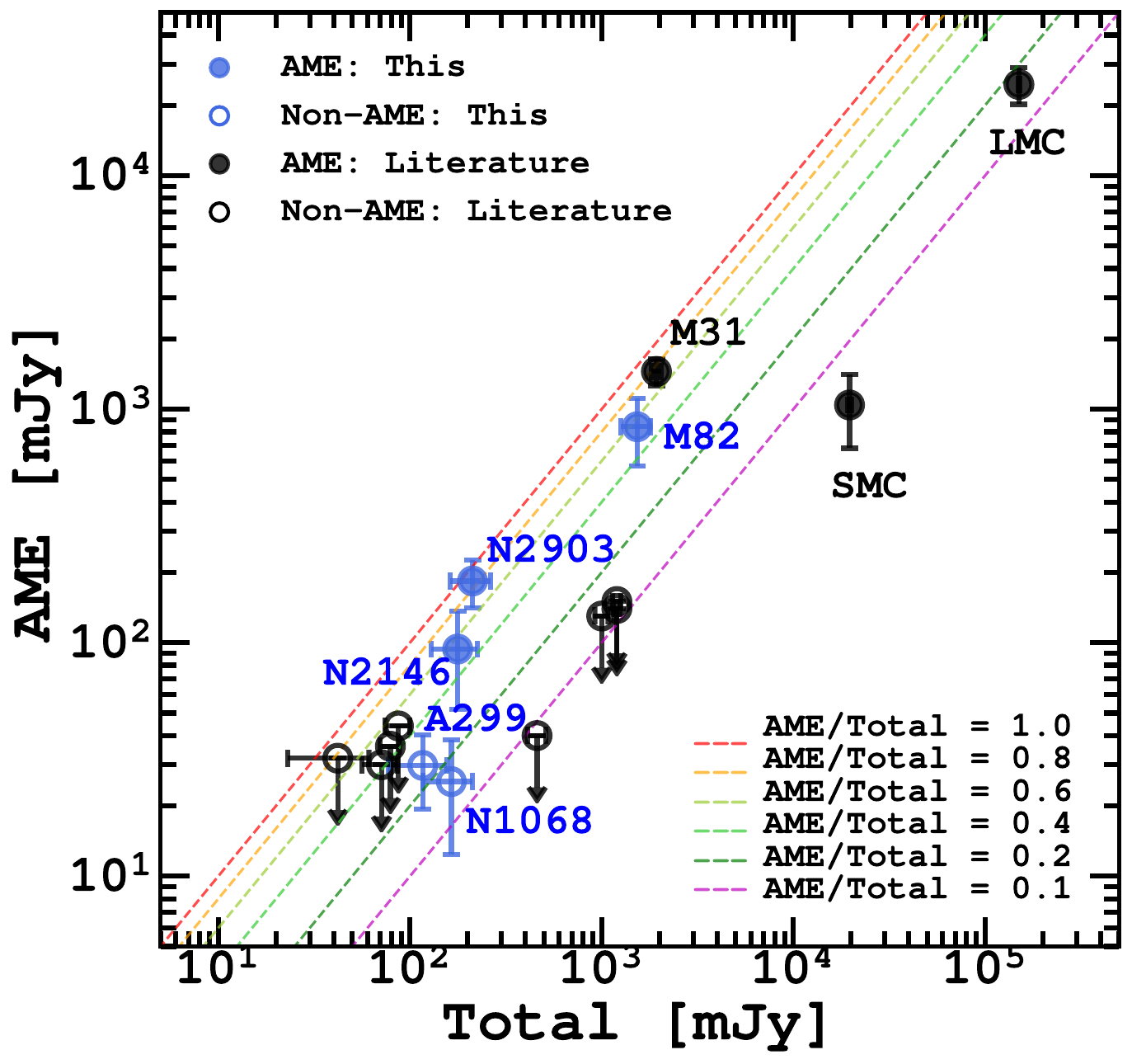}{} 
	\caption{Flux density of AME at its peak frequency, $\nu_{\rm peak}$, plotted against the total flux density at the same frequency. Dashed lines represent the constant AME/Total emission ratios, where colors correspond to different ratios varying as 0.1–1. The AME and total flux densities of our sample are measured based on our own SED while those of the previous samples are from \citet{Bianchi_2022}, who provided the measurements for their own observations as well as a collection from other literature \citep{Planck_2016, Battistelli_2019, Peel_2011, Tibbs_2018}.} 
	\label{fig:ame_total}
\end{figure}

For NGC 2903, NGC 2146, and M82, the observed data exhibited better agreement with the SED that included a bump in the range of 10–100 GHz, both visually and based on quantitative statistical evaluation, which is suggestive of AME. Meanwhile, the other two targets better matched the model with no AME bump. In Figure \ref{fig:ame_total}, the flux densities of AME at $\nu_{\rm peak}$ are plotted against the total flux density at $\nu_{\rm peak}$. The dashed lines indicate the constant ratios of AME to total emission. Even for the targets less likely to contain AME, the AME component embedded in the observed SED was still found statistically. However, for such cases (the targets on the right hand side of Fig. \ref{fig:sed}), the AME peak and total flux densities can be regarded as the upper limit (the red shaded area and the red solid line) and $\nu_{\rm peak}$ means of the frequency where the upper limit of AME peak flux density was found. 

NGC 2903 exhibited relatively large contributions of AME with $\sim70-80\%$ of the total flux at $\nu_{\rm peak}$, which is comparable to that reported in M31. In contrast, the AME component in NGC 2146 and M82 was found to contribute less, with $<50\%$ of the total strength at the peak frequency. However, our measurement for M82 was five times stronger than the upper limit previously estimated for this target by \citet{Peel_2011}.

\subsection{AME correlation with thermal dust}\label{subsec:ame_thd}
\begin{figure}
	\centering 
	\includegraphics[width=0.8\linewidth]{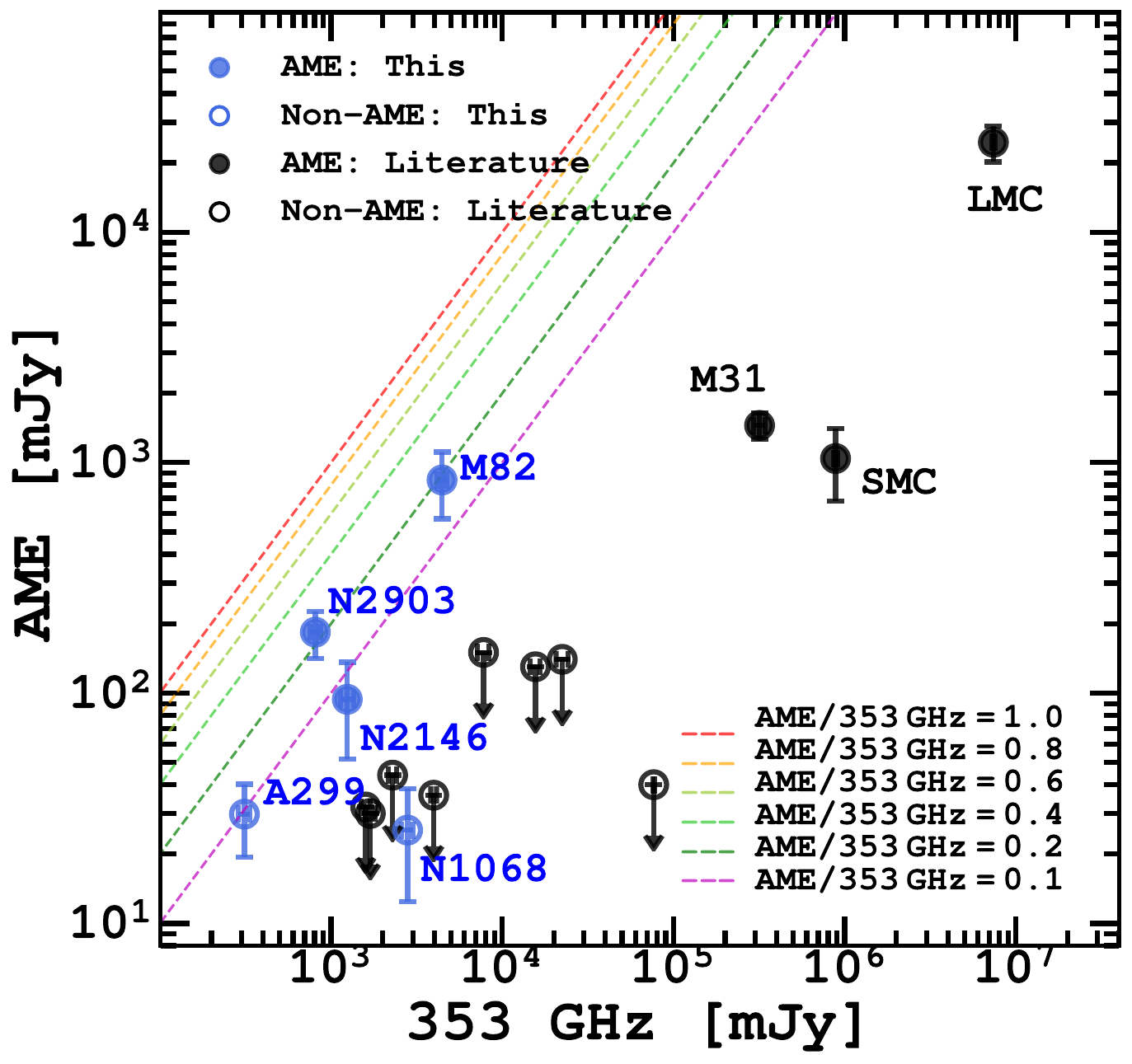}{} 
	\caption{Flux density of AME at its peak frequency, $\nu_{\rm peak}$, plotted against the thermal dust flux density at 353 GHz. Dashed lines represent constant AME/353 GHz emission ratios. The references for previous AME flux studies are the same as those in Figure \ref{fig:ame_total}. For the 353 GHz flux density, we collected data from multiple sources: our targets from our SEDs; LMC, and SMC from the SED by \citet{Planck_2016}; M31 from the SED by \citet{Battistelli_2019}; M33 from the SED by \citet{Tibbs_2018}; M82, NGC 253 and NGC 4945 from the SEDs by \citet{Peel_2011}; and NGC 3627, NGC 4254, NGC 4736, and NGC 5055 from the DustPedia database \citep{2017PASP..129d4102D}. The references of the remaining samples well align with the citation in \citet{Bianchi_2022}.}
	\label{fig:ame_353ghz}
\end{figure}

To investigate the origin of AME in our target sources and its potential correlation with dust properties, we plotted the AME flux density at the peak frequency as a function of thermal dust flux evaluated at 353 GHz, as shown in Figure \ref{fig:ame_353ghz}. Owing to the limited resolution, we conducted global comparisons among galaxies deriving both AME and 353 GHz emissions based on the SED of each galaxy. For the comparison sample, we used the AME flux densities from the data reported by \citet{Bianchi_2022}; the data comprised measurements from the various literatures as well as their own observations. The emissions at 353 GHz were derived in the same manner as that for our sample from the same references as in the study by \citet{Bianchi_2022}, that is, using the SED of individual targets and directly obtaining the value of some galaxies from the database \citep{2017PASP..129d4102D}. The plot revealed a notable trend wherein galaxies with stronger thermal dust emissions tended to exhibit more robust AME emissions, suggesting a correlation between AME and thermal dust emission. However, this result can be biased owing to the limited sample size of AME detections.

\subsection{Correlation of AME with radiation field or dust temperature}\label{subsec:ame_dusttem}
\subsubsection{AME emissivity vs. dust temperature} \label{subsubsec:emis_dusttem}
The AME emissivity at the peak frequency is dependent on the opacity and temperature of dust grains, and the relative strength of AME with respect to the thermal dust emission \citep{Bianchi_2022, 2018ARA&A..56..673G} as described below: 
\begin{equation}
\epsilon^{\rm AME}_{\nu_{\rm peak}} = \frac{S^{\rm AME}_{\nu_{\rm peak}}}{S_{3\ \rm THz}}\kappa^{\rm d}_{3\ \rm THz}B_{3\ \rm THz}(T_{\rm d})
\end{equation}
where $S^{\rm AME}_{\nu_{\rm peak}}$ is the AME flux density at the peak frequency $\nu_{\rm peak}$ of the AME component, ${S_{3\ THz}}$ is the total flux density of thermal dust emission at 3 THz (or 100 $\mu$m), $B_{3\ \rm THz}(T_{\rm d})$ is the intensity for the dust temperature $T_{\rm d}$ from the Planck function, and $\kappa^{\rm d}_{3\ \rm THz}$ is the absorption cross section per unit mass at 3 THz. For our targets, we estimated the ${S_{3\ THz}}$ flux density from our SED fit. For the comparison sample, the AME/${S_{3\ THz}}$ ratio was adopted from \citet{Bianchi_2022}, who presented these ratios based on the AME and the 3 THz flux densities from previous studies combined with their study results. By adopting the Heterogeneous dust Evolution Model for Interstellar Solids (THEMIS) \citep{2017A&A...602A..46J}, $\kappa^{\rm d}_{3\ \rm THz}$ can be estimated \citep{2018ARA&A..56..673G} as 
\begin{equation}
\kappa^{\rm d}_\nu [\rm pc^2M_{\odot}^{-1}] = 6.9\times 10^{-3}(\frac{\nu}{3\ \rm THz})^{1.79}.
\end{equation}

For the targets whose observed SED were suggestive of AME, the emissivities were estimated at $\nu_{\rm peak}$. in contrast, for the other two galaxies with no AME, the 1$\sigma$ upper limits of AME at $\nu_{\rm peak}$ from the fitting as described in \ref{subsec:sig_ame}, were adopted.

\begin{figure}[]
  \centering
  \includegraphics[width=.95\linewidth]{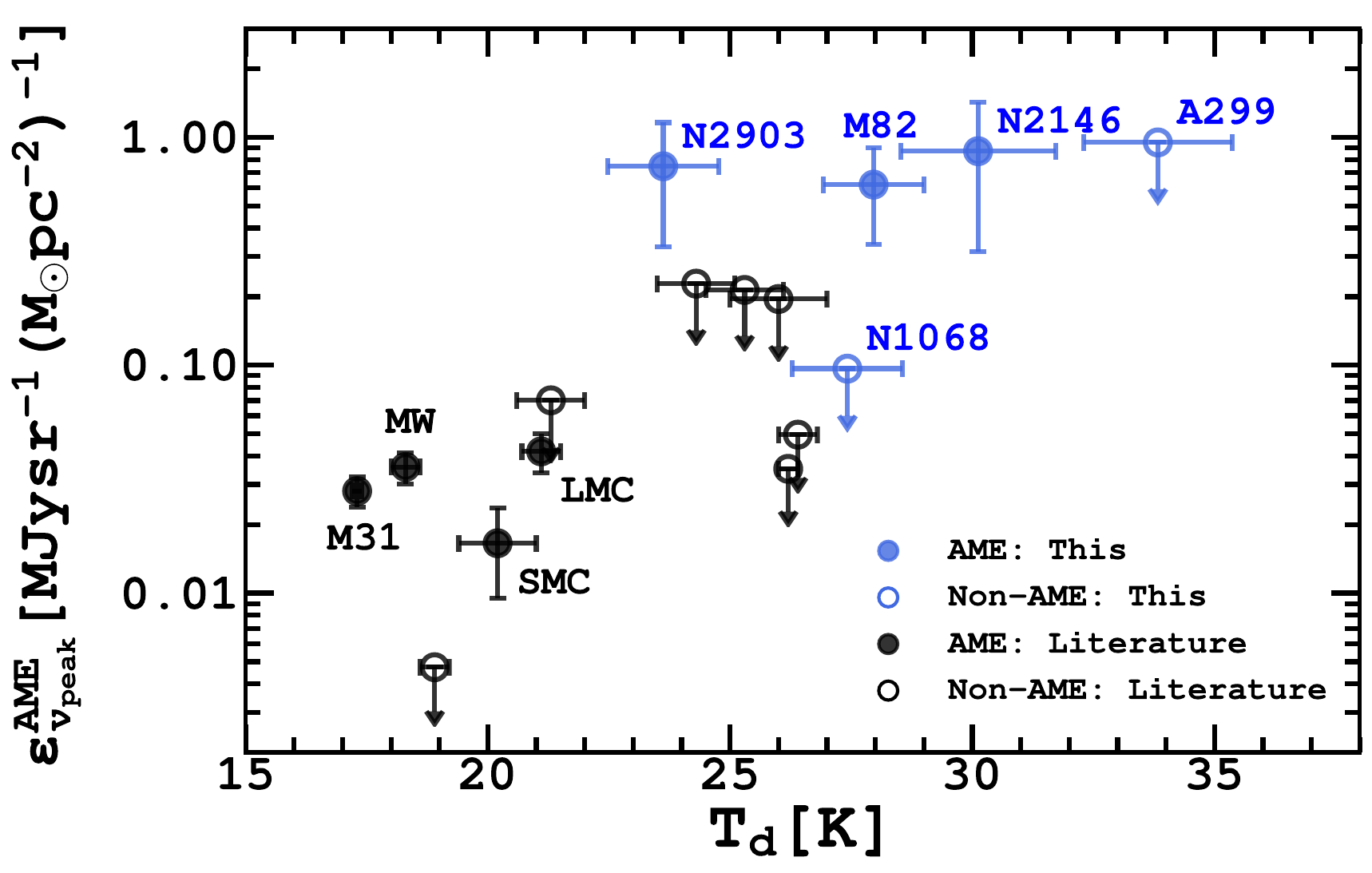}{} 
  \caption{AME emissivity vs. the dust temperature, where AME emissivities are estimated at their AME peak frequencies. The 1$\sigma$ upper limits of AME at possible peak frequency are adopted for non-detected AME targets.}
  \label{fig:emis}
\end{figure}

Figure \ref{fig:emis} shows the AME emissivities of our sample and certain other nearby targets from the literature. The AME emissivity of NGC 2903 was $\epsilon^{\rm AME}_{\nu_{\rm peak}}$ = 0.7 $\pm$ 0.4 MJy sr$^{-1}$ (M$_\odot$ pc$^{-2}$)$^{-1}$ at $\nu_{\rm peak}$ and the mean AME emissivity of three detections $\epsilon^{\rm AME}_{\nu_{\rm peak}}$ was $\approx$ 0.8 $\pm$ 0.3 MJy sr$^{-1}$ (M$_\odot$ pc$^{-2}$)$^{-1}$ at $\nu_{\rm peak}$. For comparison, the mean AME emissivity at 30 GHz for our three detections $\epsilon^{\rm AME}_{30\ \rm GHz}$ was $\approx$ 0.2 $\pm$ 0.1 MJy sr$^{-1}$ (M$_\odot$ pc$^{-2}$)$^{-1}$, which is higher by one order of magnitude than that obtained for AME detection in M31 estimated at 30 GHz ($\approx$ 0.03 $\pm$ 0.00 MJy sr$^{-1}$ (M$_\odot$ pc$^{-2}$)$^{-1}$) \citep{Battistelli_2019, Bianchi_2022}. This difference may be attributed to the environment where AME was detected as further discussed in the following section. 

Adopting the dust-to-gas ratio of 0.0074 \citep{2017A&A...602A..46J}, the AME emissivities in the unit of H-column density were 4.4 $\pm$ 2.5 $ \times 10^{-17}$ and 4.4 $\pm$ 1.5 $\times 10^{-17}$ Jy sr$ ^{-1} $ (H cm$^{-2}$)$^{-1}$ for NGC 2903 and the mean AME emissivity of three detections, respectively.  In addition, the mean AME emissivity at 30 GHz for our three detections was $ \epsilon^{\rm AME}_{30\ \rm GHz} $ $\approx$ 1.4 $\pm$ 0.6$ \times 10^{-17}$ Jy sr$ ^{-1} $ (H cm$^{-2}$)$^{-1}$, which is comparable to the predicted values by the theoretical spinning dust models for several environmental conditions such as cold neutral medium (CNM), warm ionized medium (WIM), and reflection nebulae (RN) \citep{AliHaimoud_2009}.

Interestingly, Figure \ref{fig:emis} shows that the AME emissivity tended to increase with the dust temperature, which characterized the strength of the local radiation field (denoted by $U$) because of their relationship $ T_{d}\propto U^{1/(4+\beta)}$ with $\beta$, that is, the dust opacity index \citep{2011piim.book.....D}. This trend was not expected in case of the theoretical modeling of spinning dust, which shows the weak dependence of the spinning dust emissivity on the radiation strength \citep{AliHaimoud_2009}. However, such modeling assumes that the size distribution of VSGs (e.g., PAHs) is constant for varying radiation fields. Recently, it has been suggested that in case of a strong radiation field, large dust grains could be disrupted into small and very small grains owing to centrifugal stress \citep{Hoang:2019da}. Consequently, the spinning dust emissivity increases with increasing dust temperature \citep{2020ApJ...893..138T}.

\subsubsection{AME peak frequency vs. dust temperature}\label{subsubsec:peakfreq}
From the SED fitting, the AME bumps of NGC 2146 and NGC 2903 were found to peak at $\sim$ 70-80 GHz, which is the first case identifying AME at such high frequencies to date. Meanwhile, the peak of M82’s AME was obtained at a more comparable $\nu_{\rm peak}$ with those reported both in Galactic and extragalactic objects, that is, $\sim$ 30 GHz. Thus, at least two among our three detections may originate from the region of higher temperature and/or density, whereas AME in M82 arises under similar conditions as M31 and Galactic regions or other nearby galaxies \citep[e.g.,][]{Tibbs_2012, Murphy_2010}.

The AME peak frequency has been predicted to vary with the local environments \citep{1998ApJ...508..157D, Hoang_2011}. For example, in the study of Galactic photodissociation regions (PDRs), \citet{10.1093/mnras/staa4016} and  \citet{Cepeda-Arroita_2021} found that the AME peak frequency tended to increase toward the ionizing star. This is also consistent with the models of spinning dust \citep{1998ApJ...508..157D}, which predict higher $\nu_{\rm peak}$ in the regions of stronger radiation and higher density, such as molecular cloud (MC), dark cloud (DC), reflection nebulae (RN), and photodissociation region (PDR) rather than warm ionized/neutral medium (WIM/WNM), and cold neutral medium (CNM).

\begin{figure}[] 
	\centering 
	\includegraphics[width=.92\linewidth]{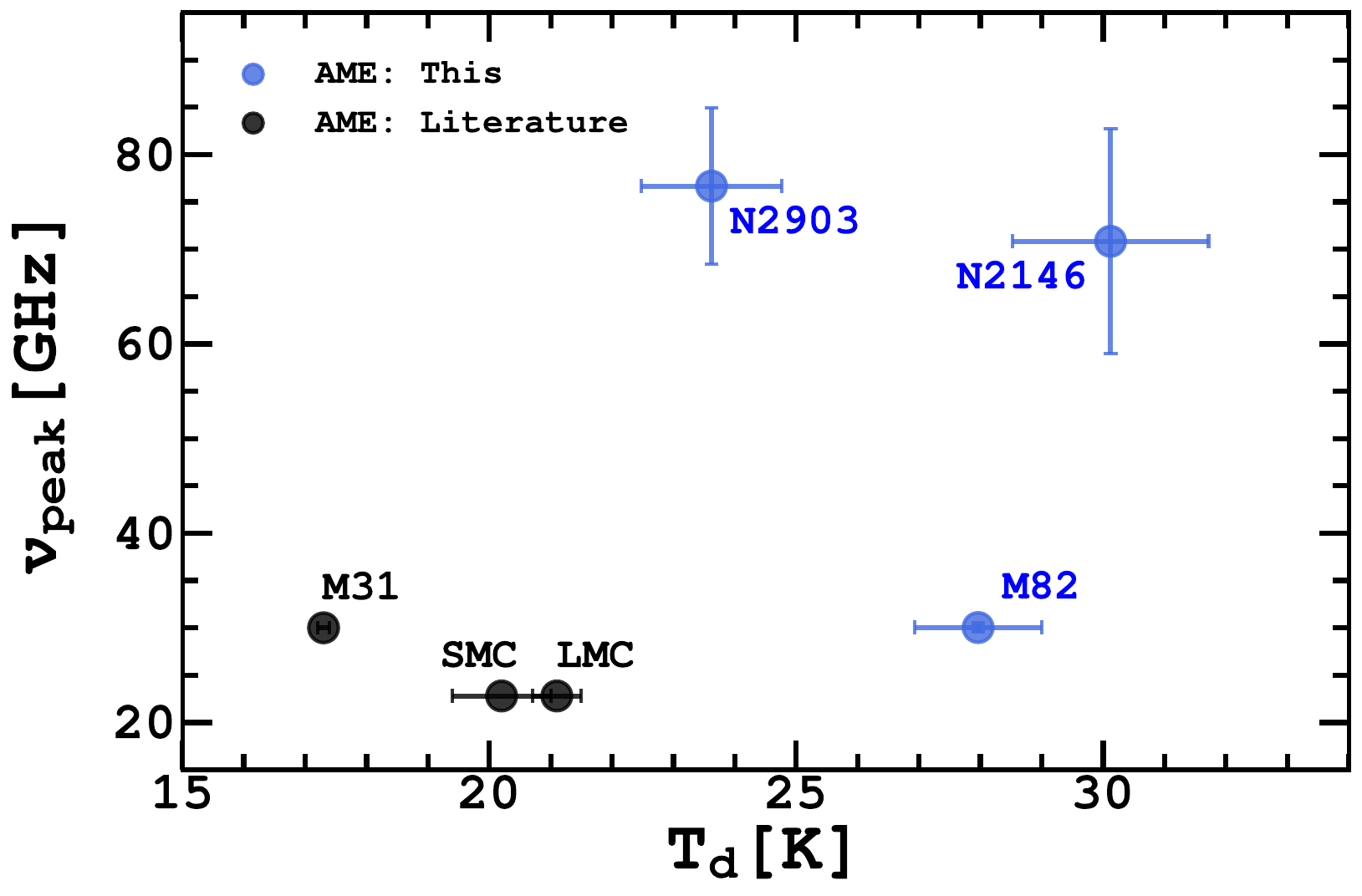} \\ 
	\caption{Correlation between the peak frequency of AME vs. the dust temperature. Our detected AME targets are shown by blue symbols while the black symbols are samples from the literature.} 
	\label{fig:Td_fpeak}
\end{figure}

Figure \ref{fig:Td_fpeak} shows the AME peak frequency versus the dust temperature from the best fit with the observed SED. Compared to the previous results, two targets of our sample (except M82) that exhibited higher $T_{\rm d}$ also indicated higher $\nu_{\rm peak}$. Thus, a certain correlation between $\nu_{\rm peak}$ and $T_{\rm d}$ was evident. 
Indeed, \cite{Cepeda-Arroita_2021} reported the increase in the AME peak frequency with the increasing dust temperature. However, M82 exhibited a higher temperature, while $\nu_{\rm peak}$ was comparable with the previous measures from the literature. This may be induced by the harsh environment of starburst systems such as M82, which affects the properties of PAHs/VSGs that emit AME.

\subsection{Correlation of AME and PAHs/small dust grains}
As AME is expected to originate from spinning dust (i.e., PAHs/VSGs), we investigate the correlation between flux densities of AME and dust tracers. We used the 8 $\mu$m and 12 $\mu$m emission as a proxy of PAHs, and the 24 $\mu$m (or 25 $\mu$m) emission as a tracer of hot VSGs. 
The Spitzer IRAC, MIPS, and IRAS data were used. The IRAC and MIPS data were corrected for the potential contamination from the stars following the prescription by \citet{Helou_2004} using 3.6 $\mu$m flux. 

\begin{figure*}[] 
	\centering 
	\includegraphics[width=.85\linewidth]{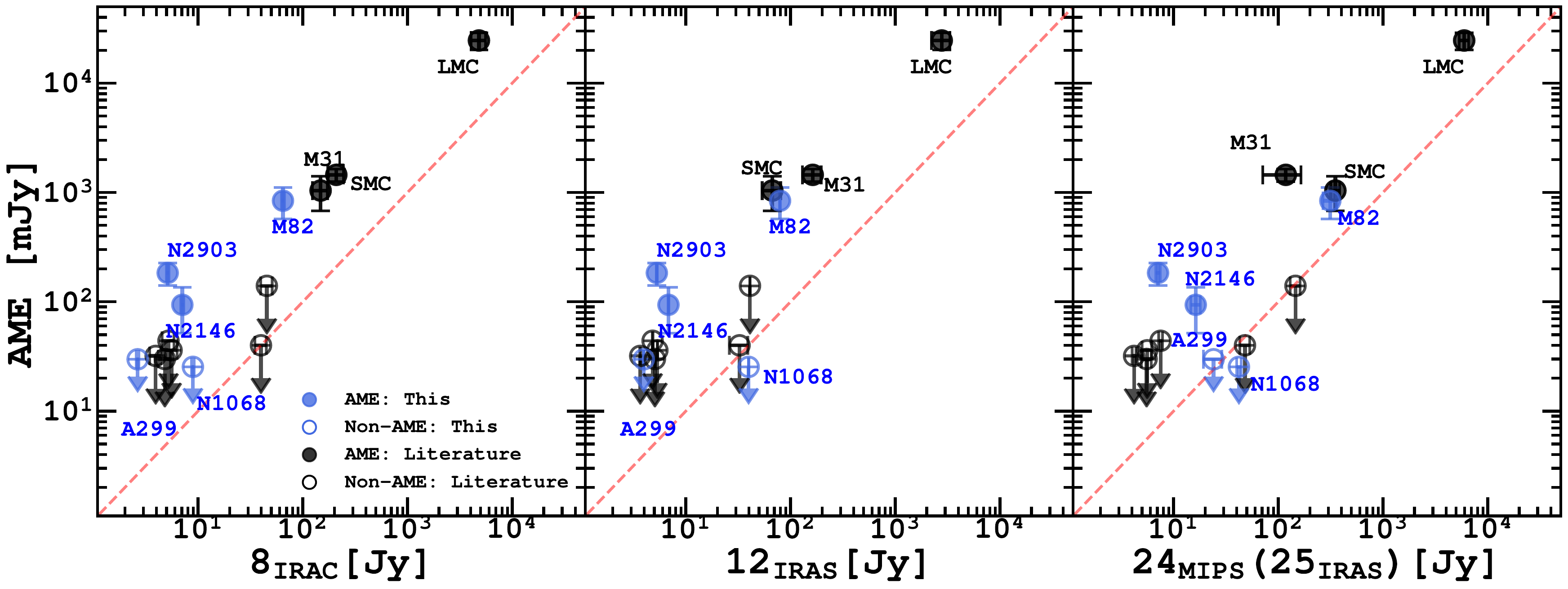} \\ 
	\caption{Correlation between the flux density of AME at $\nu_{\rm peak}$ and the 8 $\mu$m (left), 12 $\mu$m (middle) and 24 $\mu$m (right) flux densities. The 8 $\mu$m, 12 $\mu$m are PAHs tracers while the 24 $\mu$m flux density is warm dust tracers. The right plot includes the 24 $\mu$m flux densities of NGC 1068 and NGC 2903 from 24 $\mu$m MIPS data and the 25 $\mu$m flux densities of NGC 2146, M82, and Arp 299 from 25 $\mu$m IRAS data, assuming $\nu f_\nu(24\ \mu \rm m) = \nu f_\nu(25\ \mu \rm m)$. The red dashed lines are the 1:1000 constant ratio of AME to dust emission.} 
	\label{fig:ame_8_12_24}
\end{figure*}

In Figure \ref{fig:ame_8_12_24}, the correlation between AME flux density vs. the 8 $\mu$m flux density is shown in the left panel and the middle panel shows the AME flux density vs. the 12 $\mu$m flux density. These two plots were used to examine whether AME correlated to PAHs. Similarly, we plotted the AME flux density vs. the 24 $\mu$m  flux density to check whether AME correlated to the 24 $\mu$m emission emitted by very small grains. Herein, the 24 $\mu$m flux densities of NGC 1068 and NGC 2903 were measured from Spitzer MIPS data and those of NGC 2146, M82, and Arp 299 were approximated from 25 $\mu$m IRAS data, assuming $\nu f_\nu(24\ \mu \rm m) = \nu f_\nu(25\ \mu \rm m)$. This is because a saturation effect has been observed in Spitzer MIPS 24 $\mu$m data of M82 and there are no available Spitzer MIPS 24 $\mu$m data for NGC 2146 and Arp 299. In addition, we used red dashed lines to indicate the 1 mJy of AME to 1 Jy of dust emission constant ratio as a guide to the eye. In all the panels, the data points consistently followed a constant ratio (evidenced by dashed red lines). Further, an increasing trend in the flux density of AME was observed with increase in the flux density at 8 $\mu$m, 12 $\mu$m, and 24 $\mu$m. Thus, a correlation between AME and PAHs was observed, as indicated by the emission at 8 and 12 $\mu$m, as well as VSGs, which were traced by the emission at 24 $\mu$m. Although PAHs have been suggested as a source of AME \citep{2010A&A...509L...1Y, 2019PASJ...71..123B}, several studies have reported a lack of correlation between AME and PAHs, thereby ruling out PAHs as the main carriers of AME \citep{Hensley_2016, Hensley_2017, Hoang_2016_2}. We present an extended discussion of these issues in Section \ref{subsec:PAH_AME}.

\section{DISCUSSION}  \label{sec:dis}

\subsection{PAHs vs. VSGs in hot galaxies} \label{subsec:PAH_AME}
Our sample includes three starburst (SB) systems, one active galactic nucleus (AGN) host, and one SB/AGN composite as listed in Table \ref{tab:sample}. However, we did not observe AME in strong AGNs such as NGC 1068 and Arp 299, indicating that AME favored non- or rather weak-AGN hosts. This may imply that ultra-small dust grains are easily destroyed by strong AGNs. Further, this result is also consistent with the argument that PAHs, which are found to be more abundant in starburst galaxies \citep{2006ApJ...653.1129B} but absent or rare in AGNs \citep{Weedman_2005}, are the possible carrier of AME \citep{2019PASJ...71..123B}. In addition, the study by \citet{Tazaki_2020} suggested that the harsh environments of AGNs removed small silicates, which are among the potential carriers of AME.

\begin{figure*}[]
  \centering
  \begin{tabular}{c}
    \includegraphics[width=.7\linewidth]{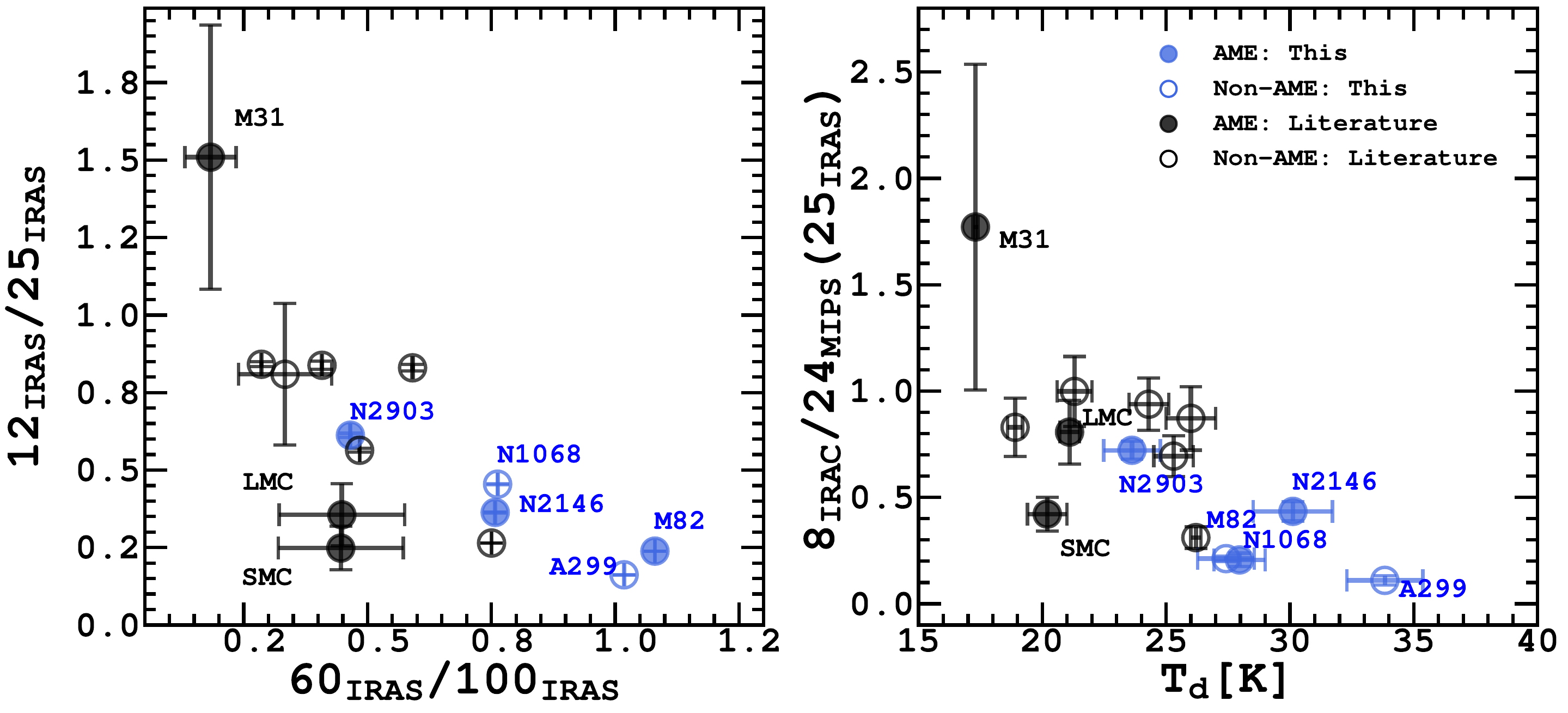}{} \\
  \end{tabular}
  \caption{Left: The ratio of 12 $\mu$m-to-25 $\mu$m tracing small dust vs. the ratio of 60 $\mu$m-to-100 $\mu$m tracing interstellar dust temperature. Right: The ratio of 8 $\mu$m-to-24 $\mu$m (or 25 $\mu$m), a proxy of the PAH fraction vs. dust temperature.}
  \label{fig:ratio_60_100_ratio_12_25}
\end{figure*}

If AME indeed originates from the spinning dust mechanism, certain correlations with the properties of VSGs are expected. To investigate the general dust properties of the sample, we plotted an IRAS color-color diagram, that is, the 12 $\mu$m-to-25 $\mu$m ratio vs. the 60 $\mu$m-to-100 $\mu$m ratio was plotted, as shown in the left panel of Figure \ref{fig:ratio_60_100_ratio_12_25}; the right panel shows the ratio of 8 $\mu$m-to-24 $\mu$m as a function of the dust temperature.

In general, in HII regions under strong radiation fields, small PAHs can be more easily destroyed \citep{1986ApJ...311L..33H, 10.1093/mnras/258.4.841}, resulting in small 12 $\mu$m-to-25 $\mu$m ratios and large 60 $\mu$m-to-100 $\mu$m ratios as observed in the sample of starburst galaxies \citep{2008A&A...484..631V}. Meanwhile, higher 12 $\mu$m-to-25 $\mu$m ratios owing to strong 12 $\mu$m radiation possibly indicate the domination of PAHs and/or VSGs. PAHs can also be destroyed by shocks from AGN feedback \citep[e.g.,][]{Zhang_2022}. However, spinning dust can also be enhanced in shocks owing to the spin-up of PAHs/VSGs by stochastic mechanical torques arising from the gas-dust drift \citep{Hoang_2019, Yoon_2022}. Indeed, the correlation between the AME and dust properties in Galactic clouds has been confirmed by \citet{2014A&A...565A.103P} such that the AME-detected regions associated with dark nebular exhibited higher 12 $\mu$m-to-25 $\mu$m ratios with small 60 $\mu$m-to-100 $\mu$m ratio. This supported the idea that AME arises from PAHs and small dust grains in the cold neutral medium (CNM) phase. 

As shown by the IRAS color-color diagram, galaxies with small ratios of 60 $\mu$m-to-100 $\mu$m exhibit higher 12 $\mu$m-to-25 $\mu$m ratio. In particular, in our sample, NGC 2146 and NGC 2903 exhibited smaller 60 $\mu$m-to-100 $\mu$m ratios with higher 12 $\mu$m-to-25 $\mu$m ratios compared to Arp 299 and M82. Thus, the fraction of smaller dust grains may be rather low in the latter two targets as in the extreme cases of starburst systems.

However, the correlation was rather weak, and the non-AME detections were essentially not distinct from the AME-detected targets. NGC 1068, wherein no AME was observed, shared similar ranges with AME-detected targets. This galaxy is a Sy2 type \citep{Kawakatu_2004} wherein the very vicinity of the central AGN is obscured, and hence possibly found with similar IR colors as the other AME detections although it has no or a very small fraction of ultra-small dust grains. The previous detections (black circles) also do not exhibit a strong correlation. SMC and LMC, which are likely to contain relatively low small-dust contents in the ISM owing to the low dust fraction, were found to be small in both ratios, and their 60 $\mu$m-to-100 $\mu$m ratios were not distinctively high compared to the rest of the sample. However, it is intriguing that M31 was found with extreme ratios, that is, the largest 12 $\mu$m-to-25 $\mu$m ratio and the smallest 60 $\mu$m-to-100 $\mu$m ratio. Among the samples shown in Figure \ref{fig:ratio_60_100_ratio_12_25}, M31 was the nearest and best representative of normal spirals where very small grains can be more sustainable following their formation. This implies that normal spirals can also contain a fair amount of ultra-small dust grains that can be detected even for distant galaxies only if allowed by the sensitivity. 

We also examined the ratio of 8 $\mu$m-to-24 $\mu$m as a function of dust temperature, as shown in the right panel of Figure \ref{fig:ratio_60_100_ratio_12_25}. The 8 $\mu$m-to-24 $\mu$m ratio can be used as a proxy of the relative PAH fraction to the total small grain content. Overall, the fraction of PAHs decreased with increase in the dust temperature; however, the correlation was rather weak. Nevertheless, as in the color-color diagram, the AME-detected targets were not clearly separable from the non-detections.

Previous studies, such as \citet{2010A&A...509L...1Y} and \citet{2019PASJ...71..123B}, have suggested a correlation between the AME and the PAH fraction at 1$^{\circ}$ angular scale (i.e., $\sim$ 2 - 40 pc) in studies of Galactic clouds. In contrast, there are several studies suggesting that AME did not correlate with PAHs. For example, \citet{Hensley_2016} studied AME with the properties of Galactic dust including PAHs, and found no correlation between AME and PAHs. Alternative candidates, such as nanosilicate grains with a radius of $\lesssim 10$ nm, have been proposed to explain the AME phenomenon \citep{Hoang_2016_2, Hensley_2016, Hensley_2017}. \citet{Hensley_2017} modified the SpDust model to compute emission arising from spinning silicate and/or iron nanoparticles and found that nanosilicate grains could consider the entirety of the observed AME while only a portion was accounted for by iron grains. This supported the theory that the carriers responsible for AME could be non-PAH nanoparticles. Thus, the question of whether PAHs are associated with AME remains a topic of ongoing debate, highlighting the need for further studies of the spinning dust mechanism and potential AME carriers. The behavior of AME appears to differ on global and local scales, which could indicate that AME properties depend sensitively on the local environments within individual galaxies. High angular resolution data would provide a better understanding of the nature of the AME carriers.

\subsection{AME contribution in radio emission}
Another way to validate the presence of AME involves assessing the radio emission that is not associated with free-free and/or synchrotron emissions. As in \citet{Murphy_2011, Murphy_2012}, we compared the specific star formation rate (sSFR) estimated using the radio emission with that calculated based on other indicators. We used various star formation tracers including 24 $\mu$m emission \citep{Murphy_2011}, which has been corrected for the emission associated with older stars using the 3.6 $\mu$m emission \citep{Helou_2004}. The SFR can also be traced using the total infrared luminosity, $L_{\rm IR}(8 \sim 1000\ \mu \rm m)$ \citep{Murphy_2012}, wherein the AGN contamination can be minimized by using only far-infrared luminosity, $L_{\rm FIR}(40 \sim 500\ \mu \rm m)$ \citep{Carpio_2008}. In addition, the combination of the total infrared and ultraviolet emissions can provide reliable SFR measurements \citep{Bigiel_2008}. Another alternative involves using radio wavelengths \citep[e.g.,][]{Murphy_2011}.  

\begin{figure}[]
  \centering
  \includegraphics[width=0.99\linewidth]{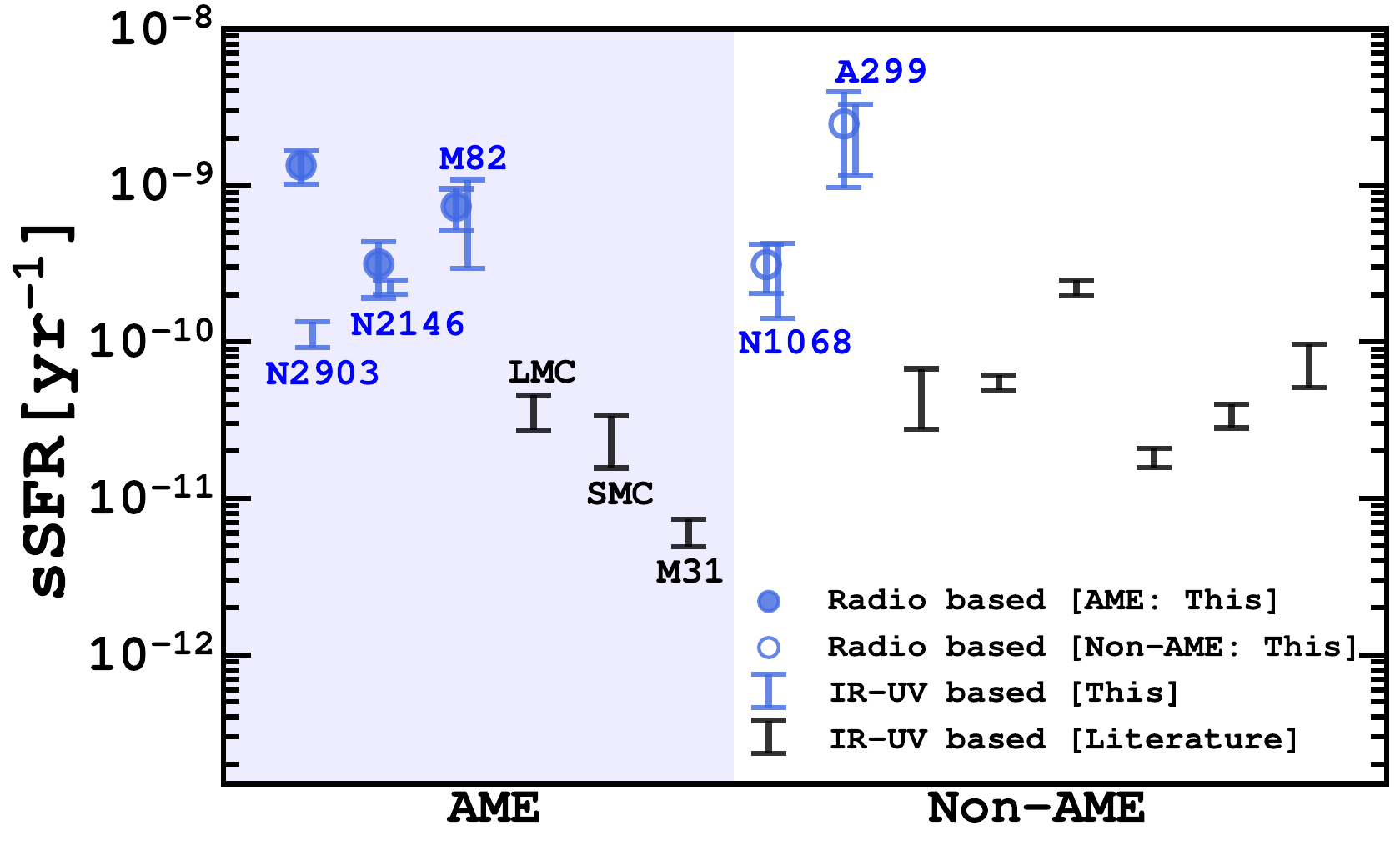}{}
  \caption{Specific star formation rates (sSFR) of the sample. The color of symbols: blue indicates our samples and black indicates samples from the literature. Only the ones with AME detection are labeled while the targets with no AME detection are not labeled. The shape of symbols: the circle shows the radio-based sSFR at $\nu_{\rm peak}$ of AME and the bar shows a range of sSFR from other tracers, including 24 $\mu$m, total IR, and FUV+IR. The shaded area highlights the AME-detected samples.}
  \label{fig:SFR}
\end{figure}

In general, the radio continuum emission comprises the thermal free–free radiation from HII regions, which is sensitive to massive stars with ages $\lesssim$ 10 Myr, and non-thermal synchrotron radiation associated with accelerated CR electrons produced in supernova remnants from the stars of $\gtrsim$ 8$\rm M_{\odot}$ with a lifetime of $\lesssim$ 30 Myr. This study used the flux measured at the peak frequency of the AME component to estimate SFR in radio. Figure \ref{fig:SFR} shows the specific star formation rate (sSFR) for our sample using different tracers. The sSFR derived from radio measurements is represented by the blue circles, and the vertical blue bars indicate the range of sSFR obtained from multiple indicators including 24 $\mu$m, total IR, and IR+FUV. In addition, the sSFR ranges from various indicators for samples from the literature are shown as vertical black bars. For comparisons, the ranges of sSFR for the AME and non-AME targets are shown in shaded areas on the left and non-shaded areas on the right of the plot, respectively. Our targets exhibited considerably higher sSFR compared to the sample from the literature owing to the selection criteria. As evident in Figure \ref{fig:SFR}, the presence of AME does not strongly depend on the star formation activity.  	

\begin{figure*}
	\centering 
	\includegraphics[width=.9\linewidth]{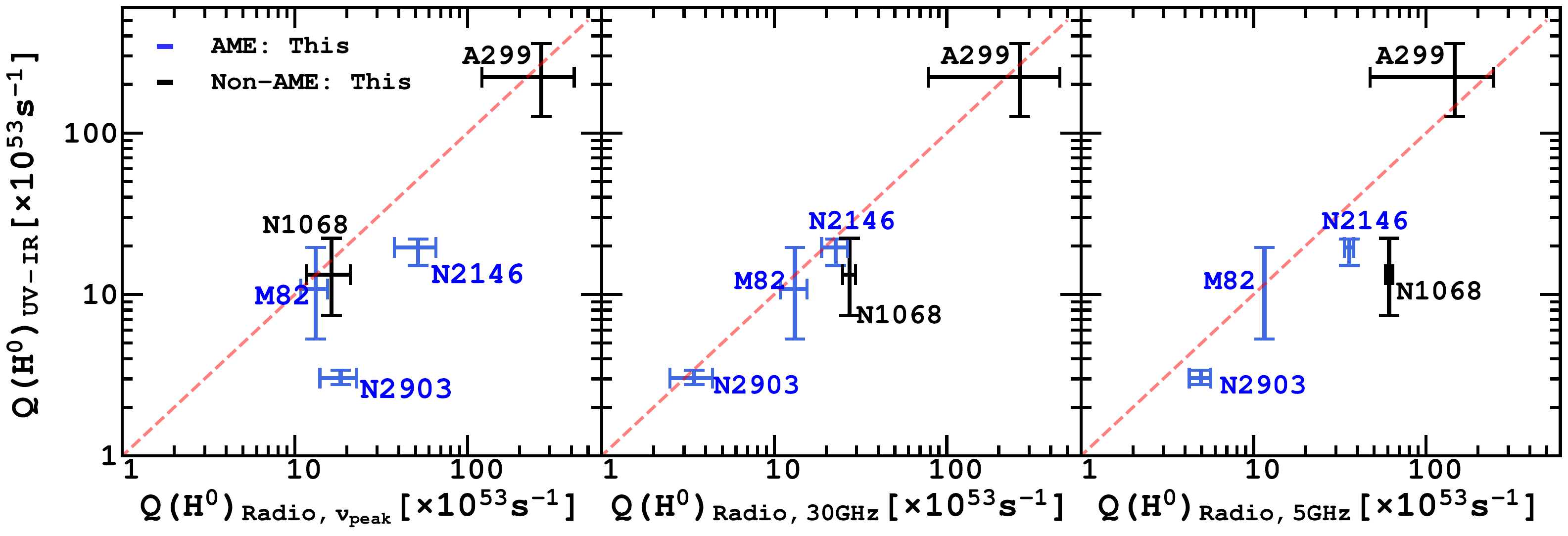} \\ 
	\caption{One-to-one correlation between the ionizing photon production rate based on radio emission and the range of the ionizing photon production rate based on multiple tracers, including 24 $\mu$m, total IR, and IR+FUV. The radio-based $Q(H^0)$ are measured at three given frequencies, including the AME $\nu_{\rm peak}$ (left panel), 5 GHz (middle panel), and 30 GHz (right panel). Blue symbols indicate our AME targets while black symbols are our non-AME targets.} 
	\label{fig: Q_ratio}
\end{figure*}

Both the optically thick free-free emission from HII regions and AME can enhance the flux in the radio regime. To identify the mechanism that contributes more, we examined the production rates of ionizing photons, $Q(H^0)$, using the IR and UV data following the relation from \citet{Murphy_2012}. Simultaneously, we estimated $Q(H^0)$ using radio emissions, assuming that they solely originated from HII regions \citep{Murphy_2010, Murphy_2011, Murphy_2012}. Figure \ref{fig: Q_ratio} shows one-to-one comparison between radio-based $Q(H^0)$ and $Q(H^0)$ based on other tracers, including 24 $\mu$m, total IR, and IR+FUV. The production rates based on the radio emission were measured at three different frequencies: the peak frequency of AME measured in this study (left panel); 30 GHz, at which several studies found the peak of AME \citep{DICKINSON20181} (middle panel); and 5 GHz, which is unlikely to be associated with AME (right panel). If radio emissions primarily originate from the HII region, radio-based $Q(H^0)$ should be more or less consistent with $Q(H^0)$ from other tracers. 

As shown by the left panel of Figure \ref{fig: Q_ratio}, NGC 2903 and NGC 2146 were offset from the one-to-one line, indicating that radio-based $Q(H^0)$ at $\nu_{\rm peak}$  was higher than $Q(H^0)$ by other tracers, while  other galaxies showed that $Q(H^0)$ lay on the one-to-one line. However, these deviations disappeared in the other two panels, indicating that the radio-based $Q(H^0)$ at 5 and 30 GHz were consistent with UV-IR based $Q(H^0)$ for all targets. These imply that the two objects with the clearest radio bump did indeed have certain other contributions in the radio wavelength such as AME. Moreover, these interpretations are supported by higher radio-based sSFR compared with sSFR from other tracers, as shown in Figure \ref{fig:SFR}.

Furthermore, in the right panel of Figure \ref{fig: Q_ratio}, NGC 1068 exhibited higher radio-based $Q(H^0)$ at 5 GHz. This galaxy is the host of a powerful AGN, and the strong synchrotron emission originated by the AGN is likely to enhance the low-frequency radio emission in this case. 

\subsection{Future perspective with extragalactic AME research}
PAHs and nanodusts play a key role in galaxy evolution because they govern gas heating and cooling, and chemical properties. More observations of AME (frequency range of 10-100 GHz) toward galaxies are required for understanding the nature of AME and its carrier \citep{DICKINSON20181}. Accurate characterization of AME is also vital for the accurate determination of star formation in galaxies using radio observations that are believed to be dominated by free-free and synchrotron emission. Existing ALMA band 1 ($\sim$ 35-50 GHz, \cite{Morata_2017}) and the upcoming SKA ($<$ 15 GHz) and ngVLA ($\sim$ 1.2-116 GHz, \cite{Murphy_2022}) will be crucial for AME research. In particular, the James Webb Space Telescope is an excellent tracer of PAHs and nanodust in mid-infrared (mid-IR). Combining PAH and nanodust observations in mid-IR by JWST with microwave observations by ALMA band 1 and future mid-SKA and ngVLA can constrain the true carrier of AME and facilitate the establishment of AME as an astrophysics diagnostic tool. 

\section{CONCLUSION}  \label{sec:conc}
This study reported the results of KVN single-dish observations performed to search for AME, while assuming a spinning dust model. The targets were a sub-sample of the MALATANG JCMT legacy survey including starburst and/or AGN nuclear activities. AME was observed in three out of five analyzed galaxies, including NGC 2903 with a significant confidence level while NGC 2146 and M82 yielded marginal detections with AME embedded in total radio emission in both galaxies. For non-detection galaxies, NGC 1068 and Arp 299, we also reported the upper limits for AME.

The AME detection in NGC 2903 and the AME marginal detection in NGC 2146 indicated that the spinning dust peaks in flux density units shifted to higher frequencies with stronger spinning dust emissivity compared to the results in the literature. These results supported the prediction of the spinning dust model that the peak frequency of AME can be increased when it originates from higher-density environments such as MC and PDRs. 

Furthermore, AME correlated to thermal dust at 353 GHz, with PAHs at 8 $\mu$m, 12 $\mu$m, and with small grains at 24 $\mu$m supporting the idea that AME was emitted by PAHs/very small dust grains. In addition, AME appeared to favor starburst galaxies rather than AGN hosts, which supported the idea that AME originated from PDRs associated with HII regions. However, strong AGNs instead destroyed small-sized dust. In addition, our analysis of dust properties implied that galaxies with detected AME tended to exhibit a higher abundance of ultra-small dust grains with lower dust temperatures, which may indicate that the abundance of tiny dust particles is crucial to the detection of AME in host galaxies. In the future, our conclusions will have to be verified using a larger size of the sample. More data will be also helpful to better understand both the nature and the role of AME in galaxy evolution. 

\vspace{5mm} 
We are grateful to the anonymous reviewer for their thoughtful and constructive comments, which improved the manuscript. We also thank Dr. Bumhyun Lee for his help with the initial observations. We are grateful to the staff of the KVN who helped to operate the array and to correlate the data. The KVN is a facility operated by the KASI (Korea Astronomy and Space Science Institute). The KVN observations and correlations are supported through the high-speed network connections among the KVN sites provided by the KREONET (Korea Research Environment Open NETwork), which is managed and operated by the KISTI (Korea Institute of Science and Technology Information). This research has used the NASA/IPAC Extragalactic Database (NED) and the HyperLeda database. This research also used the data obtained from facilities including VLA, WSRT, Effelsberg, GBT, ARO, JCMT, Spitzer, IRAS, and Herschel. P.P. acknowledges support by the Korean Government Scholarship Program (KGSP). A.C. acknowledges support by the National Research Foundation of Korea (NRF), grant Nos. 2022R1A2C100298212, and 2022R1A6A1A03053472. This work was also supported by National R$\&$D Program through the National Research Foundation of Korea (NRF) funded by the Korea government (Ministry of Science and ICT) (RS-2022-00197685). T.H. is supported by the National Research Foundation of Korea (NRF) grant funded by the Korean government (MSIT (No. 2019R1A2C1087045).  

\facilities{KVN, VLA, WSRT, Effelsberg, GBT, ARO, JCMT (SCUBA2), Spitzer (MIPS, IRAC), IRAS (IRIS, HIRES), Herschel (SPIRE, PACS)}

\software{CASA \citep{2007ASPC..376..127M, 2022PASP..134k4501C}, {\tt emcee} \citep{2013PASP..125..306F}, GILDAS/CLASS \citep{2005sf2a.conf..721P, 2013ascl.soft05010G}, Matplotlib \citep{4160265}, NumPy \citep{2011CSE....13b..22V}}

\bibliography{bibliography}{}
\bibliographystyle{aasjournal}



\end{document}